\documentclass[twocolumn,aps,superscriptaddress,prd,10pt,showpacs,preprintnumbers,nofootinbib]{revtex4-1}
\usepackage{graphicx,slashed,bbold,subfigure,amsmath,amssymb,enumerate}
\usepackage[utf8]{inputenc} % usually not needed (loaded by default)
\usepackage[T1]{fontenc}
\usepackage[english]{babel}
\usepackage{natbib} 
\usepackage[colorlinks=true,citecolor=blue,linkcolor=blue,urlcolor=blue]{hyperref}
\setcitestyle{square, numbers, comma, sort&compress} 
\usepackage{xcolor}

\begin{document}
  
\title{Phase diagram for strongly interacting matter in the presence of a magnetic field using the Polyakov-Nambu-Jona-Lasinio model with magnetic field dependent coupling strengths} 
\author{João Moreira}
\email{jmoreira@uc.pt}
\affiliation{CFisUC - Center for Physics of the University of Coimbra, Department of Physics, Faculty of Sciences and Technology, University of Coimbra, 3004-516 Coimbra, Portugal}
\author{Pedro Costa}
\email{pcosta@uc.pt}
\affiliation{CFisUC - Center for Physics of the University of Coimbra, Department of Physics, Faculty of Sciences and Technology, University of Coimbra, 3004-516 Coimbra, Portugal}
\author{Tulio E. Restrepo}\email{tulio.restrepo@posgrad.ufsc.br}
\affiliation{Departamento de F\'{\i}sica, Universidade Federal de Santa
  Catarina, Florian\'{o}polis, SC 88040-900, Brazil}
\affiliation{CFisUC - Center for Physics of the University of Coimbra, Department of Physics, Faculty of Sciences and Technology, University of Coimbra, 3004-516 Coimbra, Portugal}

\begin{abstract}

We study the phase diagram for strongly interacting matter using the 't Hooft determinant extended Nambu--Jona-Lasinio model with a Polyakov loop in the light and strange quark sectors (\emph{up}, \emph{down} and \emph{strange}) focusing on the effect of a magnetic field dependence of the coupling strengths of these interactions. This dependence was obtained so as to reproduce recent lattice QCD results for the magnetic field dependence of the quarks dynamical masses. 

A finite magnetic field is known to induce several additional first-order phase transition lines with the respective Critical End Points (CEP) in the temperature-quark chemical potential phase diagram when compared to the zero magnetic field case. A study of the magnetic field dependence in the range $eB=0-0.6~\mathrm{GeV}^2$ of the location of these CEPs reveals that the initial one as well as several of the new ones only survive up to a critical magnetic field. Only two remain in the upper limit of the studied magnetic field strength. A comparison of the results obtained with versions of the model with and without Polyakov loop is also done.

We also found that the inclusion of the magnetic field dependence on the coupling strengths, while not changing the qualitative features of the phase diagram, affects the location of these CEPs. The comparison of results with and without a regularization cutoff in the medium part of the integrals does not show a significant change.
\end{abstract}

\maketitle
%%%%%%%%%%%%%%%%%%%%%%%%%%%%%%%%%%%%%%%%%%%%%%%%%%%%%%%%%%%%%%%%%%%%%%%%%%%%%%%%
%%%%%%%%%%%%%%%%%%%%%%%%%%%%%%%%%%%%%%%%%%%%%%%%%%%%%%%%%%%%%%%%%%%%%%%%%%%%%%%%
\section{Introduction}

The study of the QCD phase diagram and the location of the Critical End Point (CEP) that marks the end of the first-order phase transition and the beginning of the crossover region, it is one of the most active fields of research of modern high energy physics. Relativistic heavy ion collisions carried out in the Large Hadron Collider (LHC) at CERN, in the Beam Energy Scam (BES) at the Relativistic Heavy-Ion Collider (RHIC) and at J-PARC in Japan have brought valuable information to describe the QCD phase diagram. In these experiments, different values of the baryonic chemical potential, $\mu_B$, are achieved by varying the beam energy, $\sqrt{s_{NN}}$, since the decrease of the energy beam gives as a result an increase in the baryonic chemical potential \cite{Andronic:2017pug}. For example, in recent years the BES-I program has analyzed data from heavy ion collisions in the energy range $7.7<\sqrt{s_{NN}}<200$ \cite{Adamczyk:2017iwn}. Despite all these efforts, definite conclusions regarding the existence (or not) of the CEP were not possible to obtain yet. New experiments will soon provide data for even lower beam energies in order to reach a higher baryonic chemical potential region and hopefully find the CEP position (or evidences of the eventual chiral transition boundaries) in the QCD phase diagram \cite{Yang:2021lfe,Yang:2017llt,Senger:2017nvf,Senger:2020fvj,Kekelidze:2017tgp}.    

In the theoretical side, lattice QCD (lQCD) calculations found that for zero chemical potential the deconfinement transition and the chiral symmetry restoration happen in a small region via a crossover \cite{Aoki:2006we}. The chiral restoration pseudocritical temperature $T_{pc}\sim 145-165$ MeV \cite{Bazavov:2014pvz,Borsanyi:2010bp,Borsanyi:2013bia} matches very well the chemical freeze-out temperature derived from the more energetic Pb-Pb central collisions carried out in the LHC \cite{Stachel:2013zma}. There the chemical freeze-out was estimated to occur at $T_{cf}\approx 156.5$ MeV for a baryonic chemical potential $\mu_B\approx 0.7~\mathrm{MeV}$ \cite{Andronic:2017pug}. Unfortunately, lQCD fails to give reliable results at medium and high baryonic chemical potentials since it suffers from the infamous sign problem \cite{Dumitru:2005ng} and only provides results for $\mu_B$ quite below the CEP. This is why the use of effective models is still essential to describe and understand the QCD phase diagram.

%%%%%%%%%%%%%%%%%%%%%%%%%%%%%%%%%%End Tulio intro 
%%%%%%%%%%%%%%%%%%%%%%%%%%%%%%%%%
Despite its simplicity, the Nambu--Jona-Lasinio model \cite{Nambu:1961fr,Nambu:1961tp} has long been regarded as an interesting tool in the study of the low energy regime of strongly interacting matter as it captures one of the main governing features of low energy hadron phenomenology, the spontaneous breaking of chiral symmetry and the associated dynamical generation of the quarks masses. The addition to this 4-quark interaction of a 't Hooft flavor determinant interaction, which corresponds to a 6-quark interaction in the $SU_f(3)$ scenario ($f$ denoting flavor), breaks the unwanted axial symmetry ($U_A(1)$) and in this formulation (which we will abbreviate as NJLH) the model has been extensively used (see for instance \cite{Klevansky:1992qe,Hatsuda:1994pi}).

One of the main shortcomings of the model is the absence of gluonic degrees of freedom and therefore an inability to describe the confinement/de\-con\-fi\-ne\-ment transition. As this is the other main governing feature of strongly interacting matter at low energy, attempts have been made to remedy this like considering an extension of the model to include the Polyakov loop dynamics \cite{Fukushima:2003fw,Megias:2003ui,Megias:2004hj,Roessner:2006xn}. The result is the well-known Polyakov extended Nambu-Jona-Lasinio (PNJL) model. 

%Discussions of the phase diagram for strongly interacting matter usually revolve around two different aspects, the transition between confined and deconfined matter (hadron/quark-gluon plasma) and the partial restoration of chiral simmetry (quarks with high dynamical mass/quarks with close to current mass). Both are expected to occur for sufficiently high temperature and/or baryonic chemical potential. A clear connection between the two is still ({\color{blue} missing? em vez de }) a hot topic of debate ({\color{red}?? subjectivo??}). The transition line separating the low temperature and low baryonic chemical potential region of the phase diagram, corresponding to the chirally broken/confined matter, from the partially chirally restored/deconfined matter is expected to become first-order below a certain critical temperature (or conversely above a critical baryonic chemical potential). A critical endpoint marks the onset of this transition behavior. Outside of this the transition corresponds to a crossover.

The inclusion of higher multi-quark interactions in the model can result in the appearance of a secondary first-order phase transition line. This has been reported in \cite{CamaraPereira:2020rtu} using the PNJL model extended to include chirally symmetric eight quark interactions \cite{Osipov:2005tq,Osipov:2006ns} and in \cite{Moreira:2014qna} using the NJLH model with eight-quark interactions supplemented with explicit chiral symmetry breaking interactions \cite{Osipov:2012kk,Osipov:2013fka} (those relevant in the sense of a $1/N_c$ expansion, with $N_c$ the number of colors). In both studies it was shown that is possible for the strange sector to also have a first-order phase transition, meaning that a second CEP for this sector can also exist in the phase diagram.

%The influence of magnetic fields has to be taken into account in a wide range of physical scenarios including heavy ion collisions, primordial Universe phases and astrophysical object such as compact stars. 
Strong magnetic fields can be found in various physical scenarios like magnetized neutron stars, early stages of the universe and heavy ion collisions at high energies. As a reference to guide our study one should note that some estimates point to the short lived production of magnetic fields of the order of $eB\sim 0.3~\mathrm{GeV}^2$ at LHC \cite{Skokov:2009qp}.
Then, it is of great relevance to know how external magnetic fields affect and modify the structure of the QCD phase diagram.
Unsurprisingly, the study of the magnetic fields effects on the phase diagram of strongly interacting matter has been a hot topic of research in recent years \cite{Menezes:2008qt,Costa:2015bza,Costa:2016vbb}, which resulted in the conjecture of the possible existence of several CEP's, with the respective first-order phase transitions lines, including the strange sector, driven by the presence of strong magnetic fields \cite{Costa:2013zca,Ferreira:2017wtx,Ferreira:2018pux}{\footnote{The existence of several first-order phase transitions at zero temperature in the presence of an external magnetic field for the one and two flavor NJL models was reported in \cite{Ebert:1999ht}. At finite temperatures an extensive study of the two flavor case can be seen in \cite{Denke:2013gha}.}.  LQCD has also performed several studies to evaluate the impact of external magnetic fields on chiral and confinement transitions \cite{Bruckmann:2013oba,Bali:2011qj,Bali:2012zg,Endrodi:2015oba}.

There is a wide range of fields where a better knowledge of the phase diagram is of great importance, at the same time helping in the interpretation of physical results and being constantly refined and constrained by new observations. From the theoretical side a useful synergy can be harnessed by combining results coming  from different techniques. Results coming from the more \emph{ab initio} approach of lQCD can, for instance, be incorporated into simple effective models.

The dynamical masses of the quarks dependence on the magnetic field (at vanishing temperature and quark chemical potential) resulting from lattice QCD calculations \cite{Endrodi:2019whh}, were recently used to obtain a magnetic field dependence of the coupling strengths of the 't Hooft determinant extended Nambu--Jona-Lasinio model in the light and strange quark sectors (\emph{up}, \emph{down} and \emph{strange}) \cite{Moreira:2020wau}. In that work the parameter space is first constrained at vanishing magnetic field, using the quarks dynamical masses and the meson spectra, and then, at non-vanishing magnetic field strength, the dependence of the dynamical masses of two of the quark flavors (\emph{down} and \emph{strange}) is used to fit a magnetic field dependence on the model couplings, both the four-fermion Nambu--Jona-Lasinio interaction and the six-fermion 't Hooft flavor determinant while keeping the momentum integration regularization cutoff fixed at its vanishing magnetic field value. This work followed several other works in which it was sought to describe the effect of the inverse magnetic catalysis by using  effective models \cite{Ferreira:2013tba,Ferreira:2014kpa,Farias:2016gmy}.
 
In the present paper we will focus our attention on the impact of this magnetic field dependence of the coupling strengths of the model in the temperature-quark chemical potential phase diagram at finite magnetic field. The presentation of this work is organized as follows. In Sec. \ref{sec2} we present the SU(3) Polyakov loop extended NJL model with 't Hooft flavor determinant. In Sec. \ref{sec3} we present the phase diagram in the temperature-quark chemical potential plane obtained at several different magnetic field strengths. We compare the results obtained with and without a magnetic field dependence of the coupling strengths and we also consider the impact of the removal of the momentum integration regularization cutoff in the medium parts of the integrals. Finally, in Sec. \ref{sec4} we draw our conclusions and final remarks.

\section{The model}\label{sec2}
\subsection{'t Hooft extended PNJL model}

The SU(3) version of the PNJL model is described by the Lagrangian density \cite{Fukushima:2003fw,Ratti:2005jh} %\cite{Klevansky:1992qe},
\begin{align}
\begin{split}
 \mathcal{L_{\rm PNJL}}=&\overline{\psi}_f\left[\slashed D^{\mu}-\hat{m}_c\right]\psi_f+\mathcal{L}_{\text{sym}}+\mathcal{L}_{\text{det}}\\
 &+\mathcal{U}\left(\Phi,\bar{\Phi};T\right)-\dfrac{1}{4}F_{\mu\nu}F^{\mu\nu},
 \end{split} 
 \label{lag}
\end{align}
where the quark sector is given by the SU(3) version of the NJL model, with the scalar-pseudoscalar and 't Hooft six fermions interactions described by \cite{Klevansky:1992qe},
\begin{align}
 \mathcal{L}_{\text{sym}}=&\frac{G}{2}\left[\left(\overline{\psi}_f\lambda_a \psi_f\right)^2+\left(\overline{\psi}_f i\gamma_5\lambda_a \psi_f\right)^2\right],\notag\\
 \mathcal{L}_{\text{det}}=&\kappa\left\lbrace\text{det}_f\left[\overline{\psi}_f\left(1+\gamma_5\right)\psi_f\right]\right.\notag\\
 &+\left.\text{det}_f\left[\overline{\psi}_f\left(1-\gamma_5\right)\psi_f\right]\right\rbrace,
\end{align}
where $\psi_f$ are the quark fields with $f={u,d,s}$, $\hat{m}_c=\text{diag}_f(m_u,m_d,m_s)$ is the quark current mass matrix, $\lambda_a$ are the Gell-Mann matrices and $G$ and $\kappa$ are coupling constants. 
The coupling between the magnetic field $B$ and quarks and between effective gluon fields and quarks is inside the covariant derivative $D^{\mu}=\partial^{\mu}-iq_fA^\mu_{EM}-iA^\mu$, where $q_f$ is the quark electric charge, $A^{EM}_\mu=\delta_{\mu 2}x_1B$ is a constant magnetic field, pointing in the $z$ direction and $F_{\mu\nu}=\partial_\mu A^{EM}_{\nu}-\partial_\nu A^{EM}_\mu$.
In the Polyakov gauge, the gluonic term, $A^\mu=gA^\mu_a(x)\frac{\lambda_a}{2}$ contribute only with the spatial component $A^\mu=-i\delta^0_\mu A^4$, where g is the strong coupling and $A^\mu_a(x)$ represents the SU(3) gauge fields.

The pure gauge sector is effectively describe by the potential \cite{Ratti:2006wg}
\begin{align}
 &\dfrac{\mathcal{U}\left(\Phi,\bar{\Phi};T\right)}{T^4}=-\dfrac{1}{2}b_2\left(T\right)\Phi\bar\Phi\notag\\
 &+b_4\left(T\right)\ln
\left[1-6\Phi\bar\Phi+4\left(\Phi^3+\bar\Phi^3\right)-3\left(\Phi\bar\Phi\right)^2\right],
\end{align}
which is fixed to reproduce lQCD results \cite{Ratti:2005jh} with 
\begin{align}
b_2\left(T\right)=&a_0+a_1\left(\dfrac{T_0}{T}\right)+a_2\left(\dfrac{T_0}{T}
\right)^2,\notag\\
b_4\left(T\right)=&b_4\left(\dfrac
{T_0}{T}\right)^3,
\end{align}
being the parameters of the potential $a_0=3.51$, $a_1=-2.47$, $a_2=15.22$ and $b_4=-1.750$. The value of $T_0$ in the Polyakov potential is usually fixed to 270 MeV according to the critical temperature for the deconfinement in pure gauge lattice results \cite{Kaczmarek:2002mc}. However, the presence of quarks in the system introduces quark backreactions that must be considered and thus a lower value of $T_0=215$ MeV, together with an accurate set of parameters to determine in the next section, is needed to obtain the pseudocritical temperature for the deconfinement given by lQCD, within the PNJL model ($T_d\approx170$ MeV \cite{Aoki:2009sc}). The expected value of the Polyakov loop is defined by $\Phi=\frac{1}{N_c}\left<\left<\mathcal{P}\exp i\int_0^\beta d\tau A_4(x_4,\overrightarrow x)\right>\right>$.

In the mean field approximation the effective quarks masses are given by the gap equations
\begin{align}
\label{GapEqs}
\left\{
\begin{array}[c]{c}%
M_u=m_u-2 G \langle \overline{\psi}_u\psi_u\rangle-2 \kappa \langle \overline{\psi}_d\psi_d\rangle\langle \overline{\psi}_s\psi_s\rangle \\
M_d=m_d-2 G \langle \overline{\psi}_d\psi_d\rangle-2 \kappa \langle \overline{\psi}_u\psi_u\rangle\langle \overline{\psi}_s\psi_s\rangle \\
M_s=m_s-2 G \langle \overline{\psi}_s\psi_s\rangle-2 \kappa \langle \overline{\psi}_u\psi_u\rangle\langle \overline{\psi}_d\psi_d\rangle
\end{array}
\right.
\end{align}
with the condensates given by
\begin{align}
\label{eqcondensate}
\langle\overline{\psi}_f \psi_f\rangle=-4 M_f \int\frac{\mathrm{d}^4p}{\left(2\pi\right)^4}\frac{1}{p_4^2+p^2+M_f^2}. 
\end{align}

To take into account the medium effects of finite temperature and/or chemical potential can be done in the usual way by replacing the $p_4$ integration by a summation over Matsubara frequencies,
\begin{align}
p_4 &\rightarrow \pi  T (2 n + 1) - i \mu  -A_4\nonumber\\
\int \mathrm{d}p_4 &\rightarrow  2\pi T \sum _{n=-\infty }^{+\infty }\,.
\end{align}

The inclusion of the effect of a finite magnetic field can be viewed as the substitution of the integration over transverse momentum, with respect to the direction of the magnetic field by a summation over Landau levels  (denoted by the index $m$) averaged over the spin  related index, $s$,
\begin{align}
  \int\frac{\mathrm{d}^2p_\perp}{\left(2\pi\right)^2}&
  \rightarrow \frac{2\pi\left|q\right|B}{\left(2\pi\right)^2} \frac{1}{2}\sum_{s=-1,+1}\sum_{m=0}^{+\infty},\nonumber\\
  \qquad p^2_\perp &
  \rightarrow (2 m+1-s)\left|q\right| B.
\end{align}
Here we have taken the direction of the $z$-axis as to coincide with that of the magnetic field such that $\overrightarrow{B}=B \hat{z}$.

The value of the confinement order parameters $\Phi$ and $\bar\Phi$ and that of the chiral condensates for a certain $T$, $\mu$ and $B$ are those that minimize the thermodynamic potential.

Let us note that to recover the standard NJL model, one only needs to take $A^\mu\to 0$ and $\mathcal{U}\left(\Phi,\bar{\Phi};T\right)\to 0$ in the Lagrangian density, Eq. (\ref{lag}). 

The vacuum integrals ($T=\mu=B=0$) are regularized using the three dimensional cutoff. The convergent thermomagnetic contributions are evaluated in two different ways for comparison purposes: by keeping the momentum integration cutoff also in all these integrals and by discarding the regularization cutoff. The pure magnetic contribution is done by using the magnetic field independent regularization scheme as done in \cite{Menezes:2008qt,Menezes:2009uc,Avancini:2011zz}. A variation of this approach to the pure magnetic contribution can be seen in \cite{Avancini:2020xqe}.

\subsection{Magnetic dependence on the coupling strengths}

Here we used our previous results from \cite{Moreira:2020wau}. Sumarinzing, the fitting procedure can be decomposed in two parts:
\begin{itemize}
\item at vanishing magnetic field we used the values for the dynamical masses $M_i$, $i=u,d,s$, the pion mass and the eta prime mass (see Table \ref{ParameterSetb}) to fit five of the six model parameters (the current masses, $m_i$ with $i=u,d,s$ and the coupling strengths, $G$ and $\kappa$). For the sixth parameter, the cutoff, $\Lambda$, we chose to impose a given value.
\item for finite magnetic field we keep the current masses and cutoff fixed at their vanishing magnetic field value and use the dynamical masses of both the \emph{down} and \emph{strange} quarks to generate a magnetic field dependence on the coupling strengths.
\end{itemize}

\begin{center}
\begin{table*}[t!]
\caption{
Model parameters (quark current masses, $m_u$, $m_d$ and $m_s$, interaction coupling strengths, $G$ and $\kappa$, and momentum integrals 3D regularization cutoff, $\Lambda$) and some properties of the low-lying pseudoscalar spectra  for the set that was used in the current work as well as the vacuum dynamical masses of the quarks ($M_u$, $M_d$ and $M_s$).  The fit was done in \cite{Moreira:2020wau} and reproduces the dynamical masses of the quarks reported in \cite{Endrodi:2019whh} as well as the mass of the pion ($M_{\pi^\pm}=140~\mathrm{MeV}$) and the eta prime mesons ($M_{\eta'}=958~\mathrm{MeV}$) at a vanishing magnetic field strength. For completeness the kaon mass ($M_{K^\pm}$), the decay width of the eta prime meson $\Gamma_{\eta'}$ and the eta meson mass ($M_\eta$) are also presented below (for more details see  \cite{Moreira:2020wau}). Values used for the fitting procedure are marked with $^*$.} 
\label{ParameterSetb}
\begin{footnotesize}
%\begin{tiny}
\begin{tabular*}{1\textwidth}{@{\extracolsep{\fill}}rrrrrrrr@{}}\hline
&\multicolumn{1}{c}{$m_u$ [MeV]} & \multicolumn{1}{c}{$m_d$ [MeV]} & \multicolumn{1}{c}{$m_s$ [MeV]} 
&\multicolumn{1}{c}{$G~\left[\text{GeV}^{-2}\right]$} & \multicolumn{1}{c}{$\kappa~\left[\text{GeV}^{-5}\right]$} & \multicolumn{1}{c}{$\Lambda~\left[\text{GeV}\right]$} &\\
\hline
& $5.89826$ & $6.01442$ & $180.450$ &  $9.47574$ & $-149.340$ & $0.600^*$ &\\
\hline
\hline
\multicolumn{1}{c}{$M_{\pi^\pm}$ [MeV]} & \multicolumn{1}{c}{$M_{K^\pm}$ [MeV]} & \multicolumn{1}{c}{$M_\eta$ [MeV]} 
&\multicolumn{1}{c}{$M_{\eta'}$  [MeV]} & \multicolumn{1}{c}{$\Gamma_{\eta'}$  [MeV]} & \multicolumn{1}{c}{$M_u$  [MeV]} & \multicolumn{1}{c}{$M_d$  [MeV]} & \multicolumn{1}{c}{$M_s$  [MeV]}   \\
\hline
$140^*$ & $557.196$ & $530.023$ & $958^*$ & $222.571$ &  $311.505^*$ & $311.684^*$ & $550.007^*$\\
\hline
\end{tabular*}
\end{footnotesize} 
\end{table*}
\end{center}

The dynamical masses of the quarks are displayed together with lQCD values and with the ones that would result from keeping the couplings fixed at their vanishing magnetic field values in Fig. \ref{MiFit}. Despite the poor agreement in the comparison of $M_u$ between the $G(B),~\kappa(B)$ case and lQCD values, the overall agreement is much better than the one obtained considering no magnetic field dependence in the couplings.

The magnetic field dependence of the couplings $G$ and $\kappa$ is displayed in Figs. \ref{GvsB} and \ref{KvsB}. A decrease in the strength of the former and an increase in the latter with increasing magnetic field is obtained (in $\kappa$ we are referring to the absolute value).

\begin{figure*}[!htb]
\center
\subfigure[]{\label{MiFit}\includegraphics[width=0.32\textwidth]{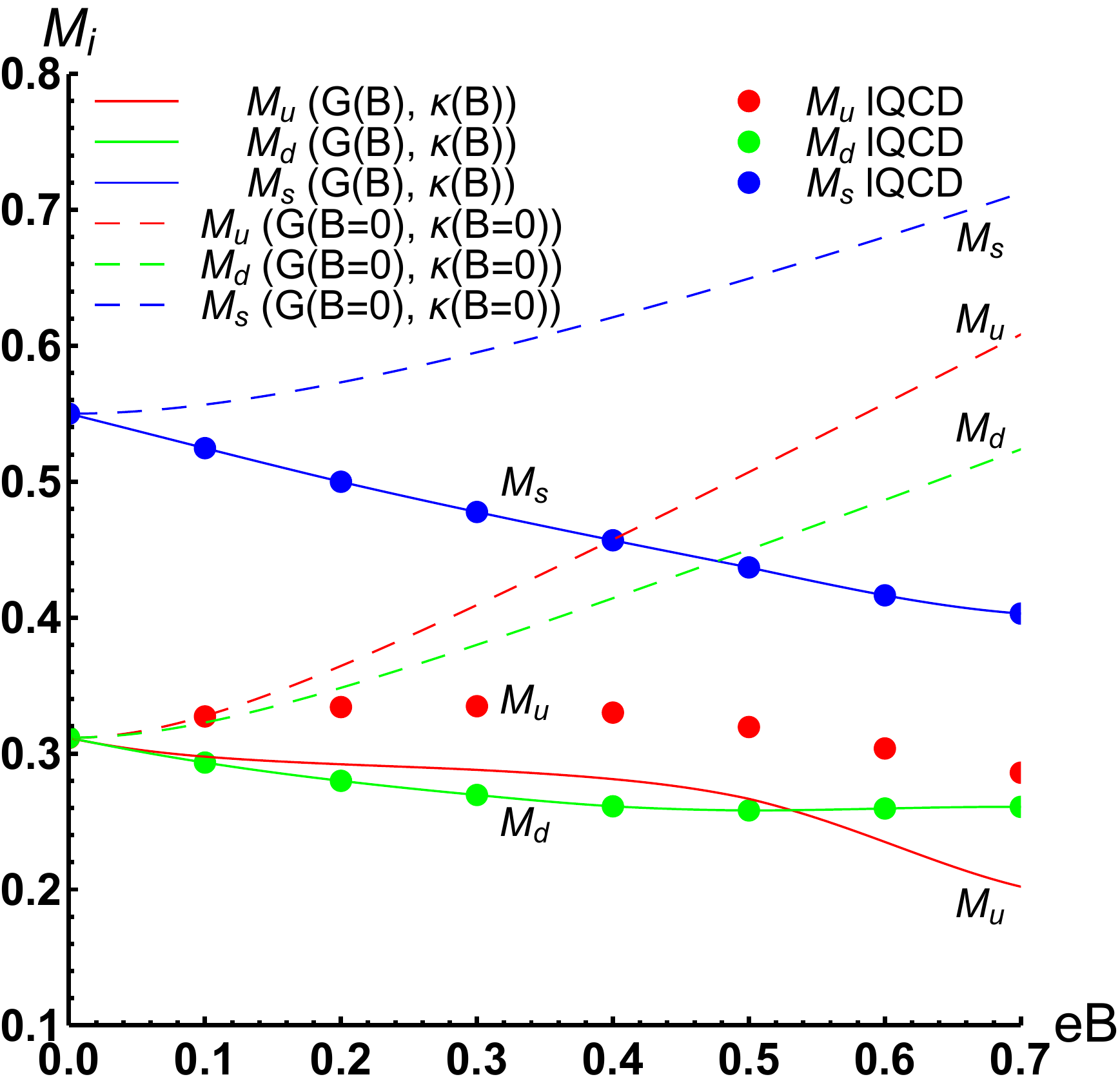}}
\subfigure[]{\label{GvsB}\includegraphics[width=0.32\textwidth]{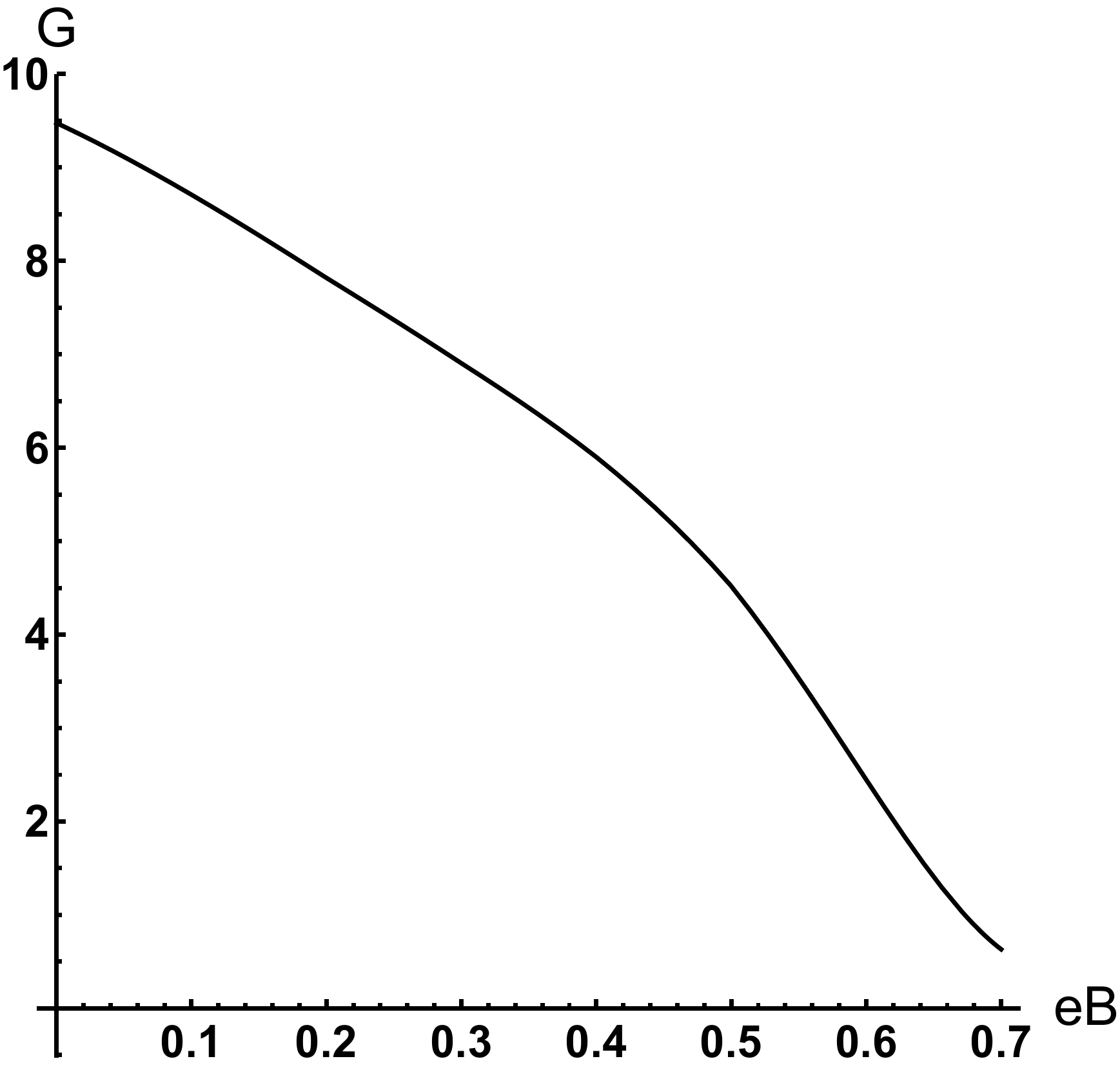}}
\subfigure[]{\label{KvsB}\includegraphics[width=0.32\textwidth]{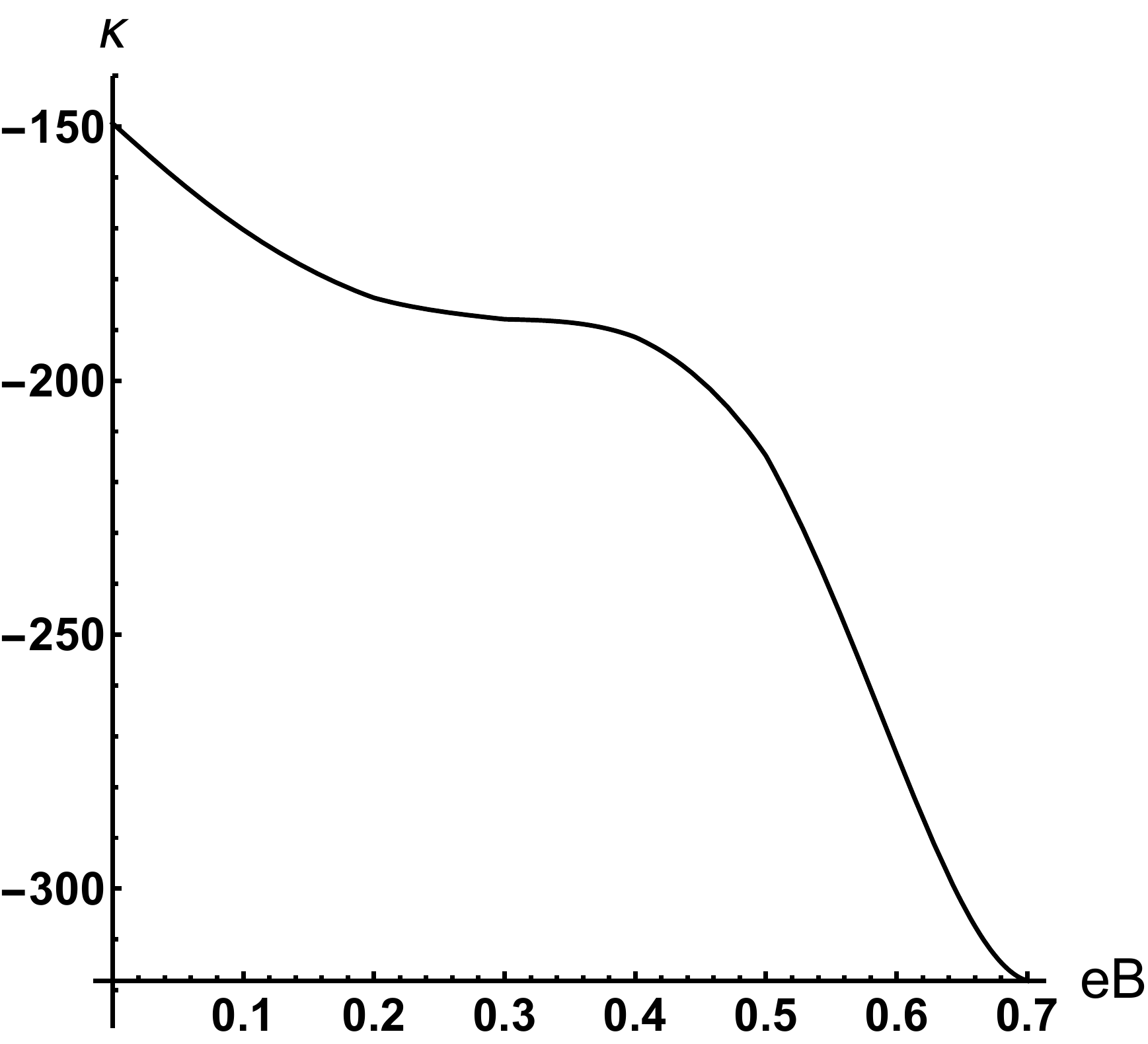}}
\caption{
The magnetic dependence of $M_d$ and $M_s$ as given by lQCD calculations \cite{Endrodi:2019whh} was used in \cite{Moreira:2020wau} to obtain a magnetic dependence on $G$ and $\kappa$. In \ref{MiFit} the dynamical masses fit and the comparison with the case where the couplings are fixed at their $B=0$ value  ($[M_i]=~\mathrm{GeV}$, $i=u,d,s$ and $[eB]=~\mathrm{GeV}^2$). In \ref{GvsB} and \ref{KvsB} the magnetic field dependence of NJL and 't Hooft interaction coupling strengths ($[G]=~\mathrm{GeV}^{-2}$ and $[\kappa]=~\mathrm{GeV}^{-5}$).
}
\label{GKFits}
\end{figure*}

\section{Results}\label{sec3}

As reported previously in \cite{Costa:2013zca,Ferreira:2017wtx,Ferreira:2018pux}, the consideration of strong external magnetic fields induces the appearance of extra first-order phase transition lines (and associated CEPs) in the phase diagram. In Fig. \ref{PD_eB_0000_0080} the phase diagram for vanishing magnetic field is shown displaying the existence of a single first-order phase transition line and the associated CEP and compared to the phase diagram at a magnetic field strength of $eB=0.080\mathrm{GeV}^2$ with its multiple transitions (in a range of magnetic fields around $0.080~\mathrm{GeV}^2$ the phase diagram can present a maximum of seven first-order lines as can be seen in Fig. \ref{PD_eB_0080} and that we will discuss later). Dashed lines denote the results obtained with removal of the cutoff in the medium contributions to the integrals. As can seen in Fig. \ref{PD_eB_0000} this pushes the CEP towards lower quark chemical potential. As for quark chemical potentials lower than the cutoff (or more precisely $\sqrt{\mu_q^2-M^2}<\Lambda$, with $M$ the dynamical mass of the quarks and $\mu_q$ the quark chemical potential) this removal must be irrelevant at vanishing temperature (only states up to the Fermi energy are occupied) it is unsurprising that the deviation is limited to higher temperatures. In fact for the shorted lines the deviation is negligible. For the remainder of the results presented here we will keep the momentum integration cutoff in all integrals, a choice which has been shown to provide more thermodynamical consistent results \cite{Moreira:2010bx}.
%\footnote{{\color{violet}For more discussions about this topic see Refs. \cite{Fukushima:2003fw,Bratovic:2012qs,Avancini:2020xqe}}}.

\begin{figure*}[!htb]
\center
\subfigure[]{\label{PD_eB_0000}\includegraphics[width=0.34\textwidth]{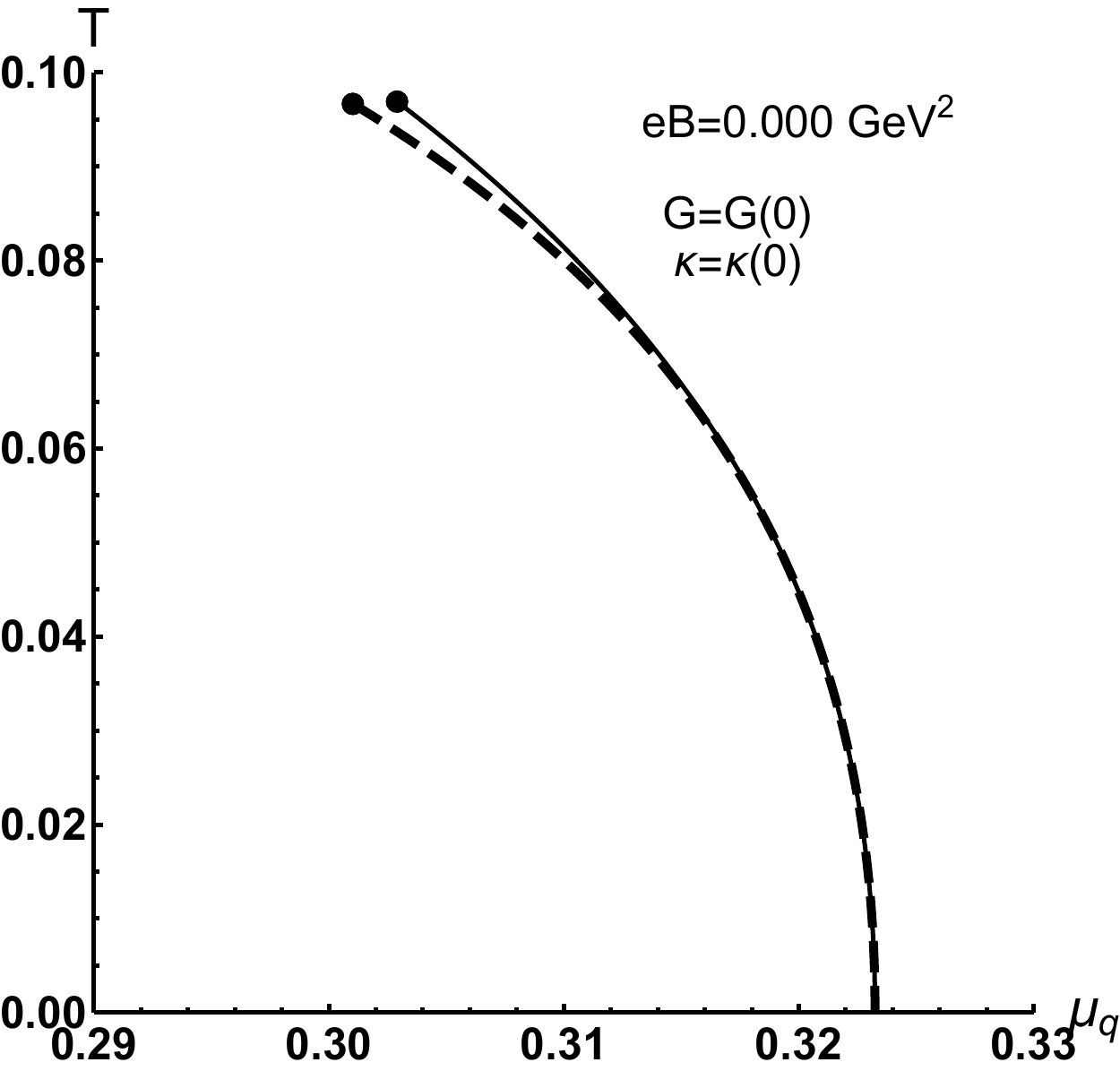}}
\subfigure[]{\label{PD_eB_0080}\includegraphics[width=0.34\textwidth]{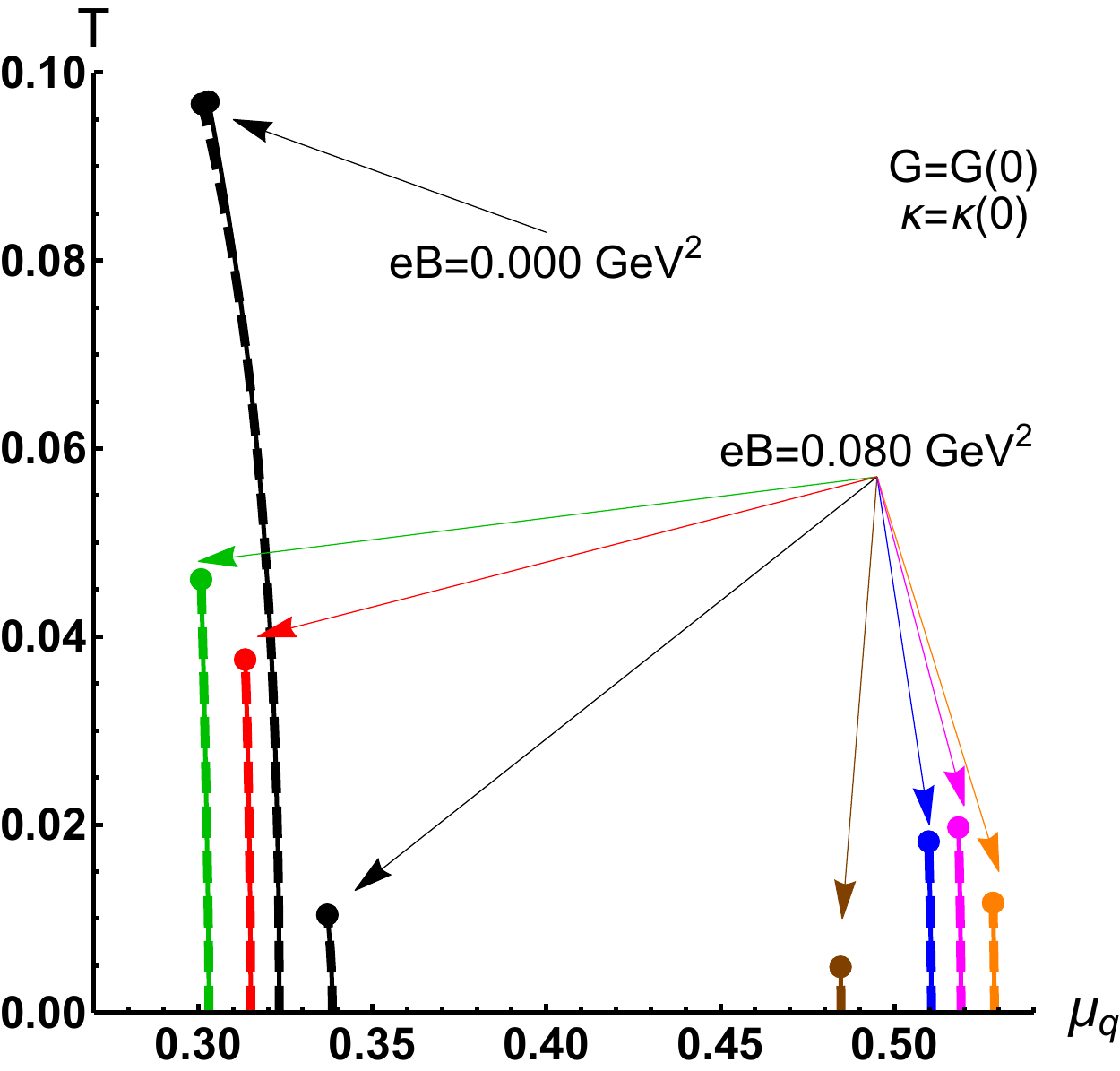}}
\caption{
Phase diagram in the temperature-quark chemical potential plane ($[T]=[\mu_q]=\mathrm{GeV}$). In the left-hand side panel for vanishing magnetic field strength ($eB=0$) displaying a single CEP and in the right-hand side panel an example of a more complex phase diagram that is obtained at finite magnetic field. Dashed lines denote the removal of the momentum integration cutoff in the medium parts as applied in the regularization of the integrals. No magnetic field dependence on the coupling strengths is considered here.
}
\label{PD_eB_0000_0080}
\end{figure*}
  
In Fig. \ref{GKBFits} we compare the phase diagrams obtained by considering the coupling strengths of the interactions determined at vanishing magnetic field versus the magnetic field dependent value. We display zooms of the temperature-quark chemical potential phase diagram around the three first-order phase transitions which are found to exist at $eB=0.300~\mathrm{GeV}^2$. We consider four scenarios: $G$ and $\kappa$ magnetic field dependent, $G$ kept fixed at the $B=0$ value with magnetic field dependent $\kappa$, magnetic field dependent $G$ and $\kappa$ at $B=0$ value and finally both couplings kept at the respective $B=0$ value.

The weakening of the NJL interaction induced by the magnetic field results in shorter first-order phase transition lines as well as their shift towards lower chemical potentials as can be seen in the shift when considering the $G(0),\kappa(0)\leftrightarrow G(B),\kappa(0)$ and the $G(0),\kappa(B)\rightarrow G(B),\kappa(B)$ cases. Both these effects are expected as a direct result of the smaller dynamical chiral symmetry breaking effects.

The increase of the flavor mixing 't Hooft determinant relevance (it increases in magnitude with the magnetic field) results in a shift of the transition starting at the lowest chemical potential at vanishing temperature (the leftmost panel, Fig. \ref{FitComparisonB0300I}) towards higher temperature and quark chemical potential as can be seen by comparing the $G(0),\kappa(0)\leftrightarrow G(0),\kappa(B)$ and the $G(B),\kappa(0)\leftrightarrow G(B),\kappa(B)$ cases. This can be a result of the stronger interplay between the light and the strange sectors. For the other transitions (middle and rightmost panel, Figs. \ref{FitComparisonB0300II} and \ref{FitComparisonB0300III}) there is no discernible effect which could be explained by the fact the light sector is already restored in that region of the phase diagram and as such an increase of the flavor mixing has no impact on those lines.

\begin{figure*}[!htb]
\center
\subfigure[]{\label{FitComparisonB0300I}\includegraphics[width=0.32\textwidth]{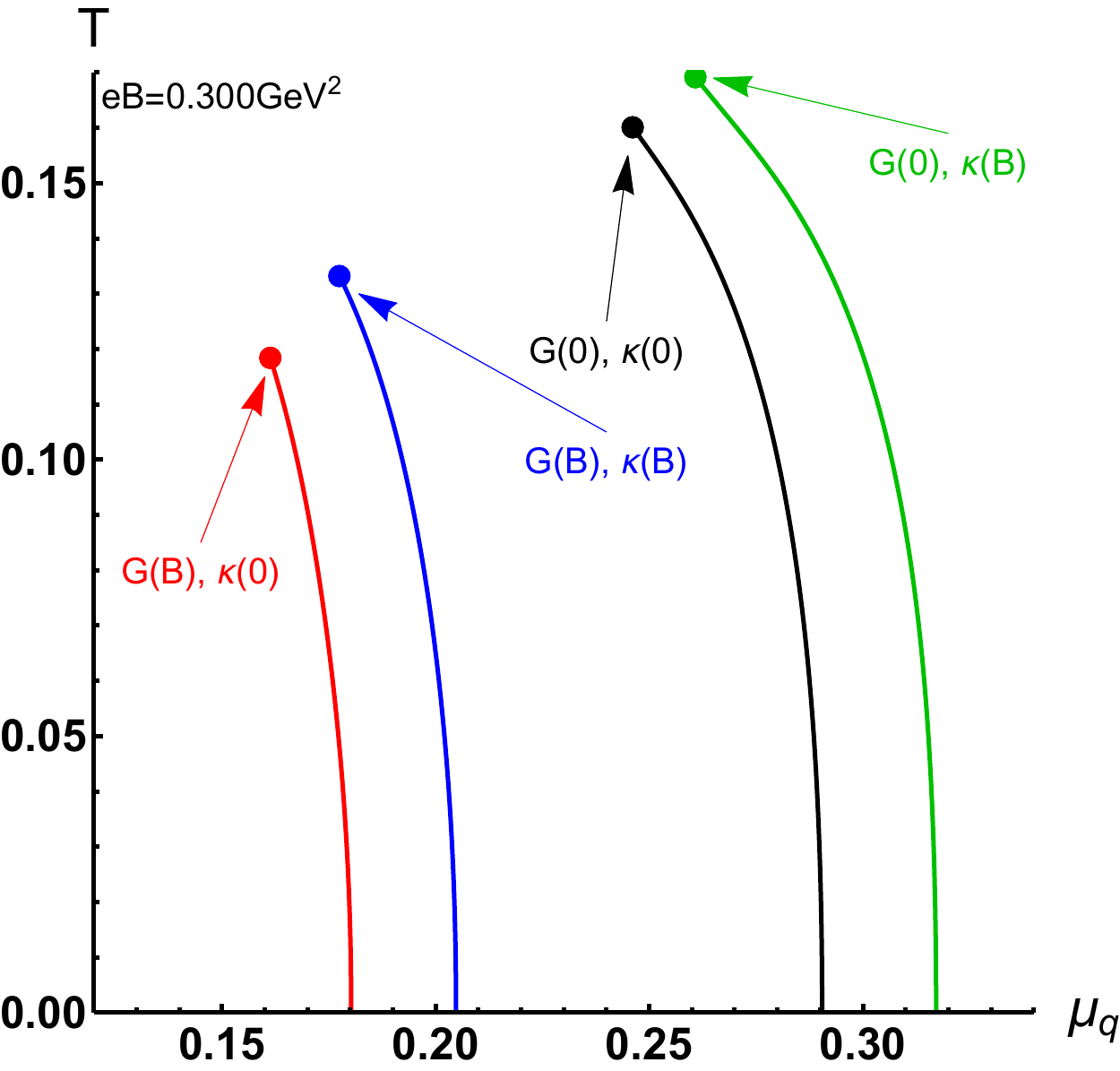}}
\subfigure[]{\label{FitComparisonB0300II}\includegraphics[width=0.32\textwidth]{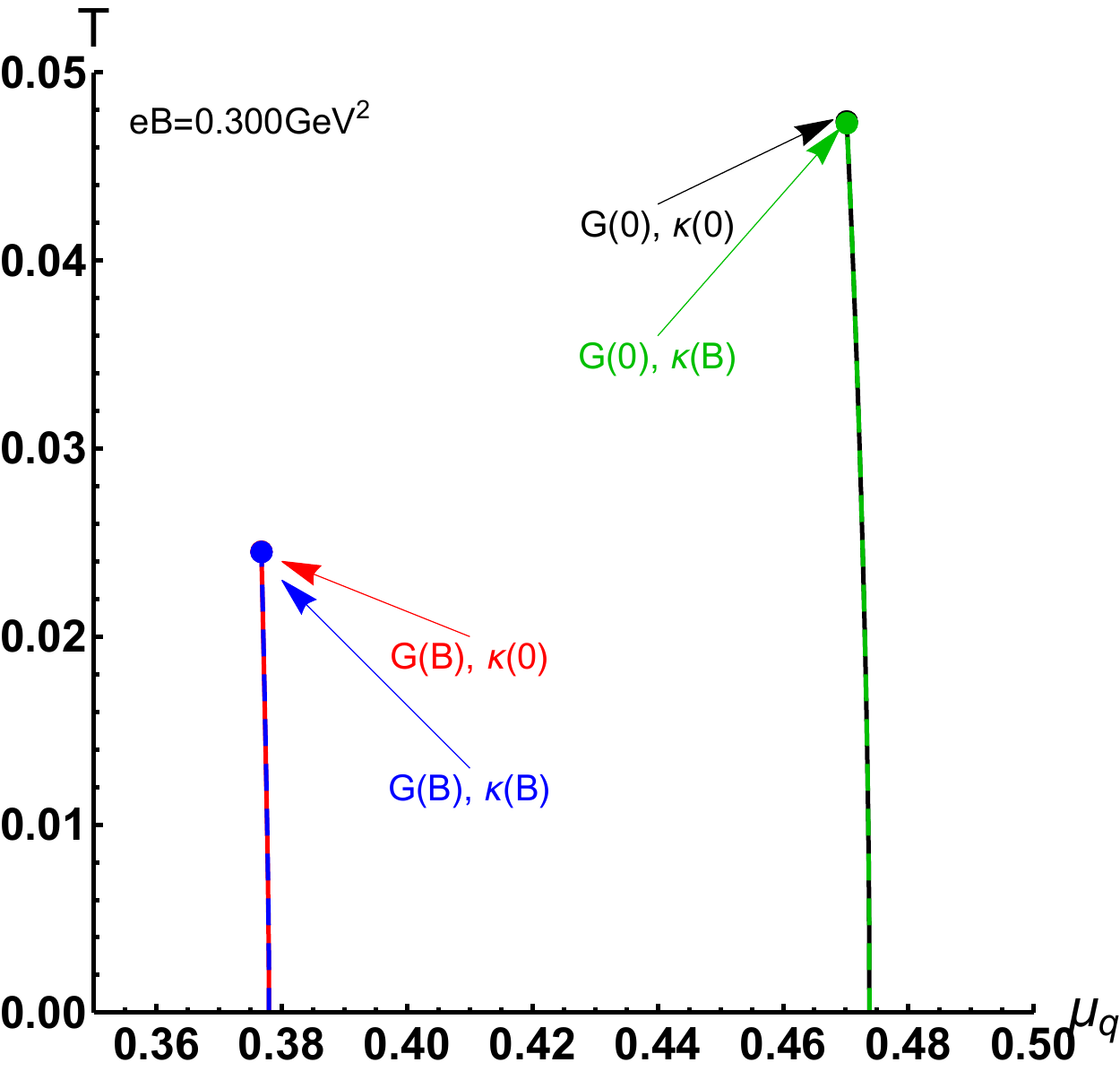}}
\subfigure[]{\label{FitComparisonB0300III}\includegraphics[width=0.32\textwidth]{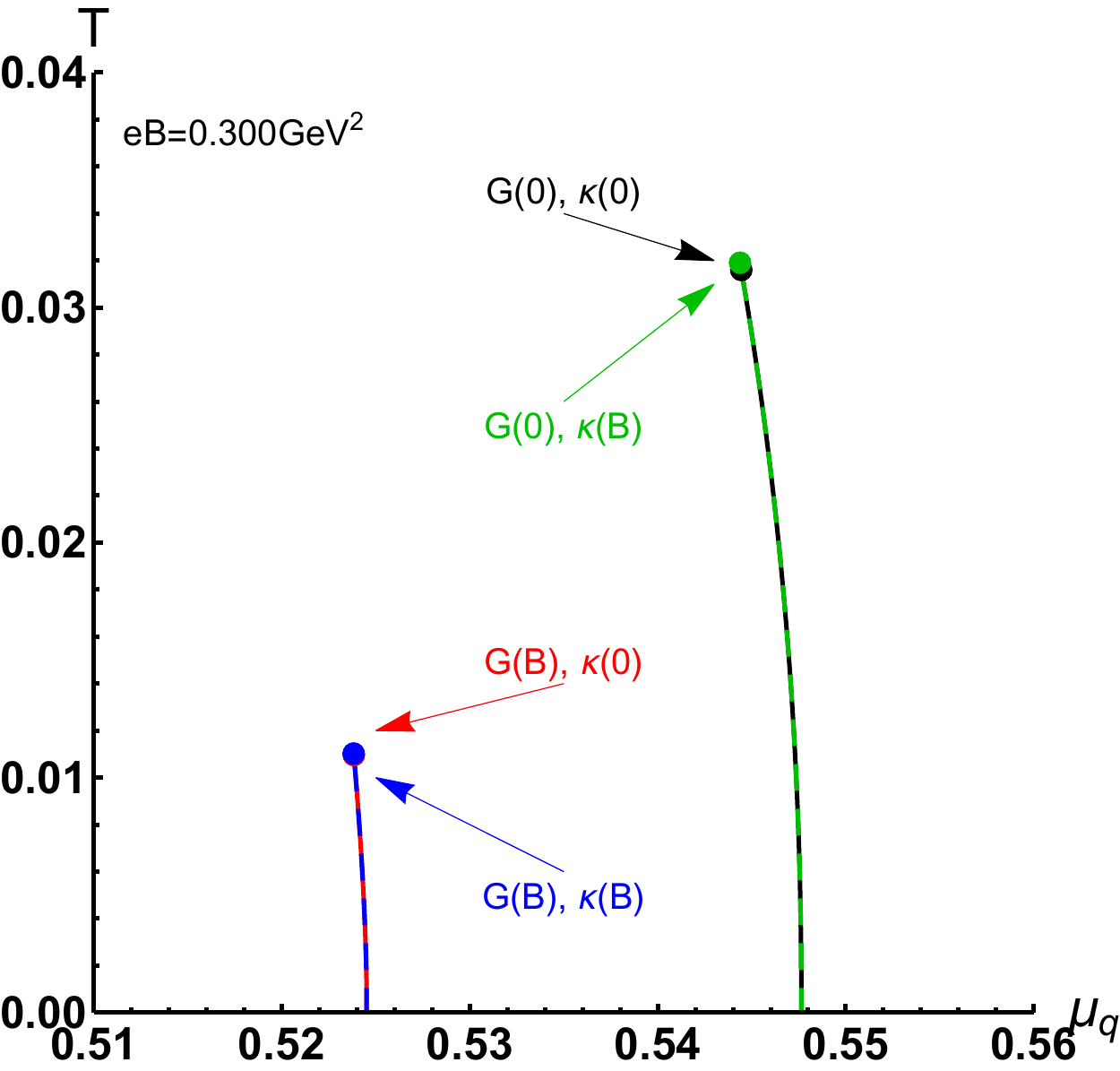}}
\caption{
Phase diagram in the temperature-baryonic chemical potential plane ($[T]=[\mu_q]=\mathrm{GeV}$) for a magnetic field strength of $eB=0.300~\mathrm{GeV^2}$. The three figures depict zooms around the three first-order phase transition lines present and show the impact of the parametrization choice by displaying four scenarios: $G$ and $\kappa$ fixed at the vanishing magnetic field values (black lines), $G$ kept  fixed at the vanishing magnetic field value but with a magnetic field dependent $\kappa$ (red lines), magnetic field dependent $G$ and $\kappa$ kept at the vanishing magnetic field value (green lines) and finally a magnetic field dependent $G$ and $\kappa$.
}
\label{GKBFits}
\end{figure*}

The impact of the external magnetic field in the $\mu-T$ phase diagram can be seen by observing the magnetic field dependence on the position of the CEPs as seen in Fig. \ref{CEPsvseB}. First of all, it is important to note that in this implementation of the extension to include Polyakov loop dynamics that sector becomes irrelevant for zero temperature (the PNJL model reverts to the NJL model). As the transition lines for the studied cases all end in a CEP (none of them spans the whole phase diagram reaching zero chemical potential), the number of CEPs is the same for both the NJL and the PNJL cases. The critical values of the magnetic field strength which correspond to the appearance or disappearance of a CEP must also be the same in both (looking at Figs. \ref{TvseBG0K0} and \ref{TvseB} we are referring to the point where the lines touch the magnetic field strength axis). The CEP temperatures obtained in the PNJL variant of the model are, as usual, higher than those obtained in the NJL model. As expected, only in the longer lines (those ending in a higher temperature CEP) we do get a noticeable change in the CEP's quark chemical potential (PNJL obtained lines end at a higher temperature/lower chemical potential). 

The number of CEPs present in the phase diagram depends on the external magnetic field strength, starting with a single CEP case at vanishing magnetic field (which we will label as $I$) located at
\begin{align}
\{\mu_q,T\}^{NJL}_{I}|_{B=0}=\{308,40\}\nonumber\\
\{\mu_q,T\}^{PNJL}_{I}|_{B=0}=\{303,97\}  
\end{align}
($[\mu_q]=[T]=~\mathrm{MeV}$).
Corresponding to the end of the usual first-order phase transition line associated with partial chiral restoration in the light sector, this CEP disappears with increasing magnetic field strength (see the line labeled as $I$ in Fig. \ref{CEPsvseB}). For small finite magnetic field the largest jump is in the down quark dynamical mass. 

With an increase in the magnetic field strength six additional CEPs eventually appear for a critical magnetic field strength $eB\sim 0.01~\mathrm{GeV}^2$ which we label from $II$ to $VII$. The initial CEP (the one labeled as $I$) disappears at a critical magnetic field strength $eB\sim 0.1~\mathrm{GeV}^2$. The phase diagram reaches a maximum complexity (in terms of number of transitions) in this magnetic field range with seven CEPs simultaneously present in the $\mu-T$ plane (recall Fig. \ref{PD_eB_0080} for the phase diagram at $eB=0.080~\mathrm{GeV}^2$). 
Some of these only survive up until a certain critical value of the studied magnetic field strengths (one must also recall the existence of a regularization cutoff which sets a validity scale on the model). For high enough magnetic field only two are left (labeled $II$ and $III$) at the upper limit of the studied range of $eB=0-0.6~\mathrm{GeV}^2$. It should be pointed out that for very low temperatures (approximately below $1~\mathrm{MeV}$) an accurate determination of the position of CEPs becomes extremely difficult. At vanishing temperature and/or with a slightly different parametrization of the quark interaction part, the occurrence of a few additional first-order phase transitions with CEPs with a very low temperature is not to be excluded.

CEPs labeled as $I$, $II$ and $V$ correspond to partial chiral symmetry restorations in the light sector whereas the remaining  refer to partial chiral symmetry restorations in the strange sector. The corresponding chemical potentials are accordingly grouped within two very narrow ranges for magnetic fields close to the critical magnetic field for their appearance but their separation increases with increasing magnetic field. The CEPs corresponding to the strange sector transitions have a much higher associated quark chemical potential.

\begin{figure*}[!htb]
\center
\subfigure[]{\label{TvseBG0K0}\includegraphics[width=0.49\textwidth]{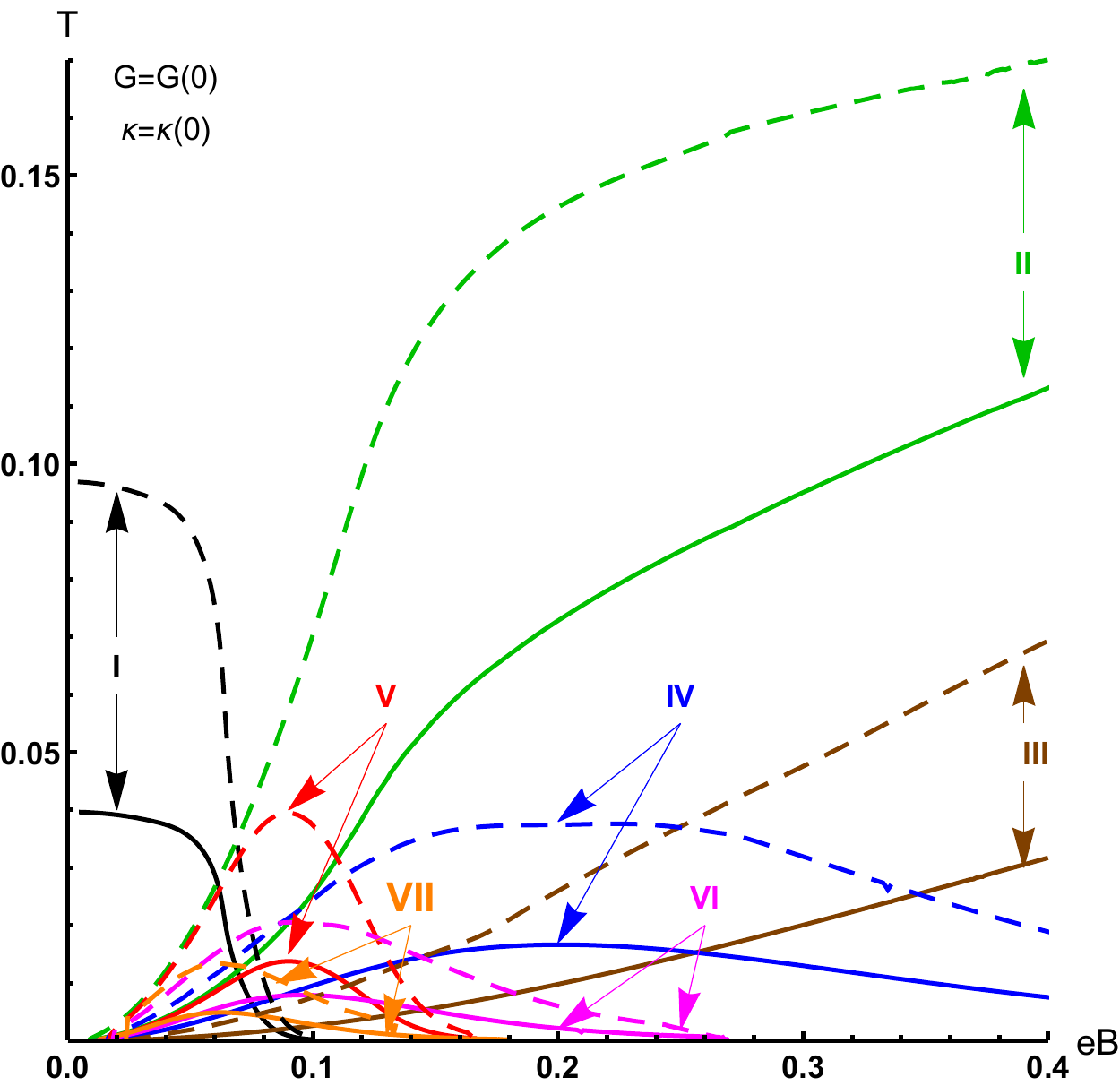}}
\subfigure[]{\label{muvseBG0K0}\includegraphics[width=0.49\textwidth]{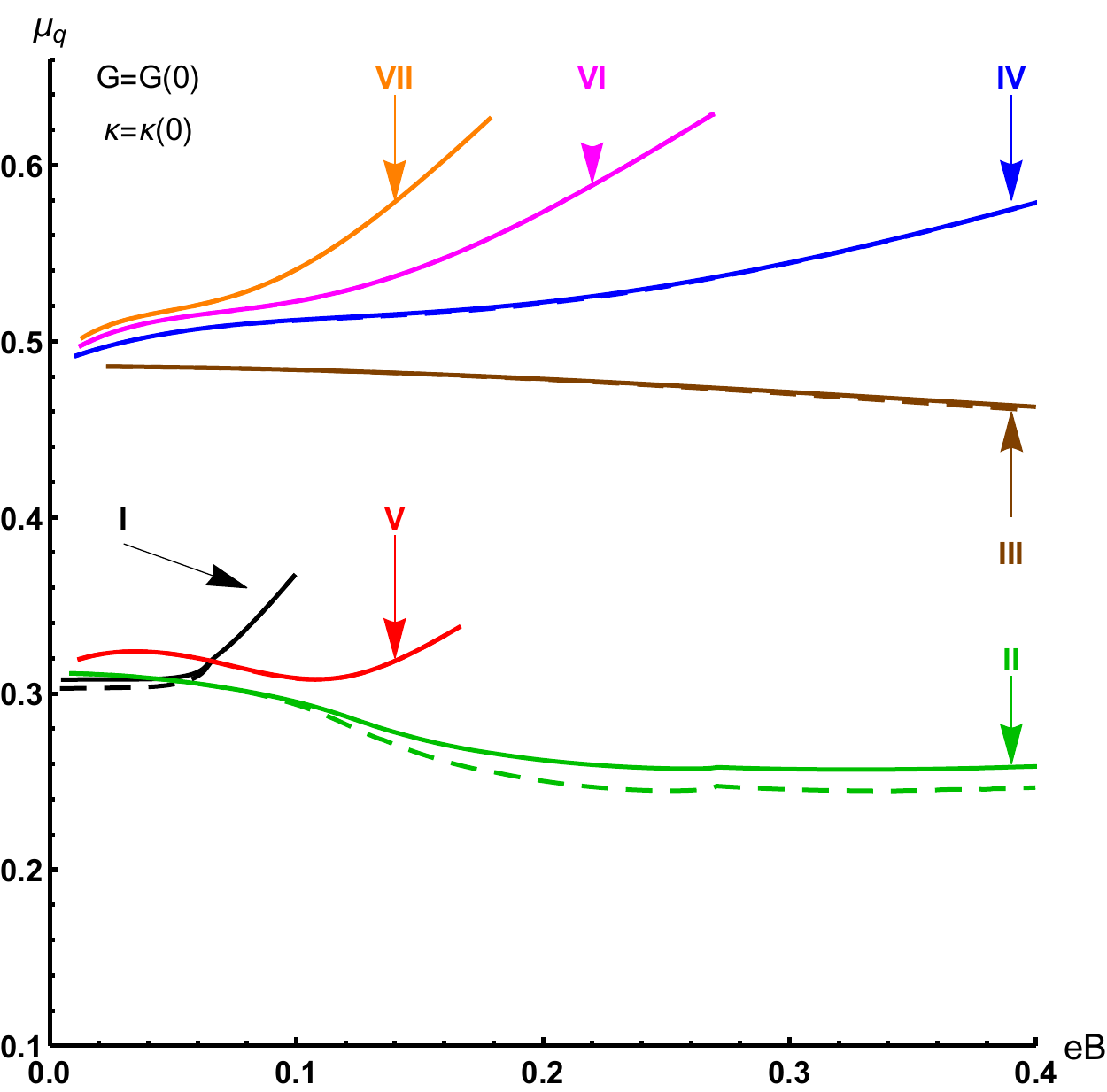}}\\
\subfigure[]{\label{TvseB}\includegraphics[width=0.49\textwidth]{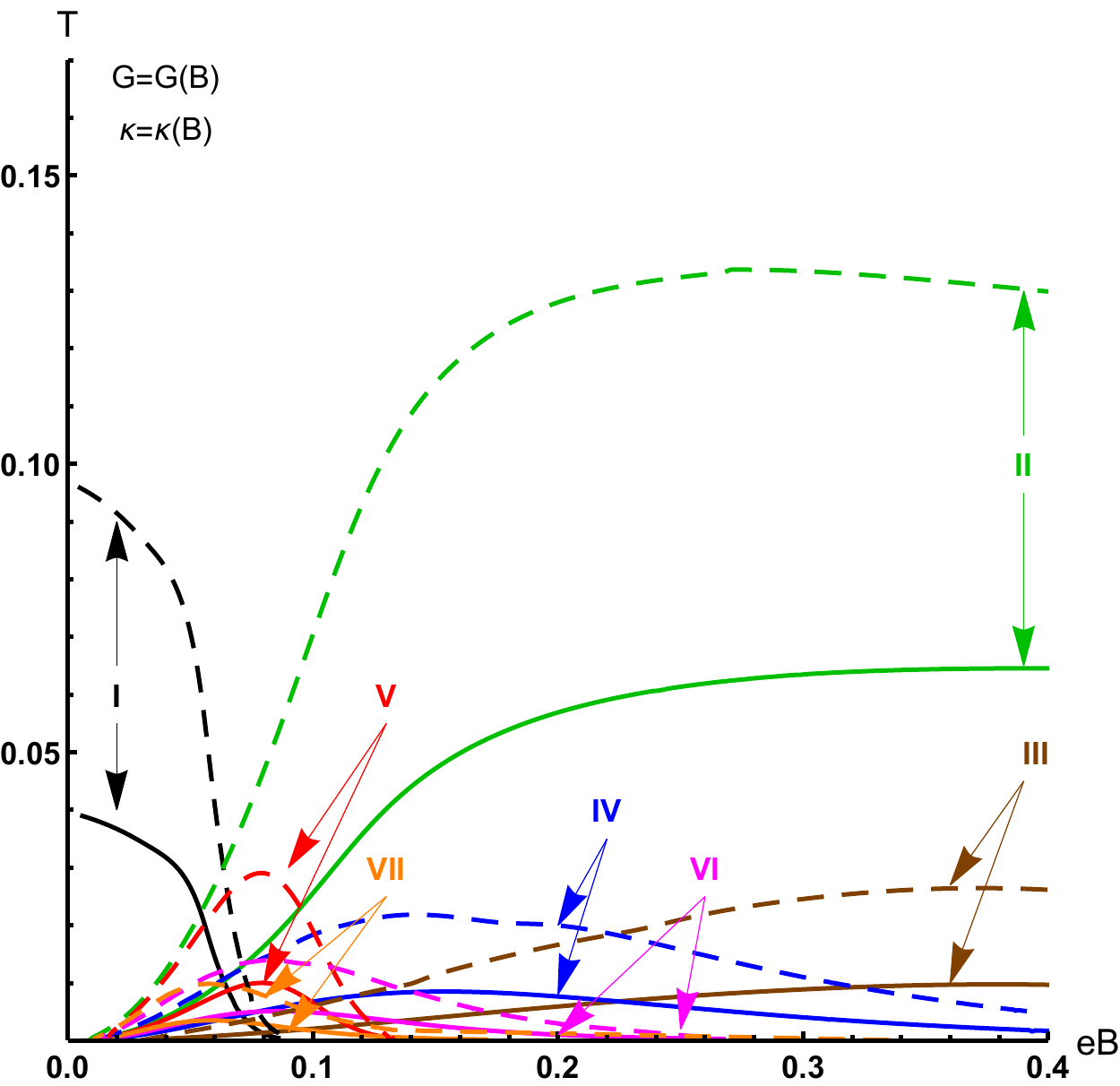}}
\subfigure[]{\label{muvseB}\includegraphics[width=0.49\textwidth]{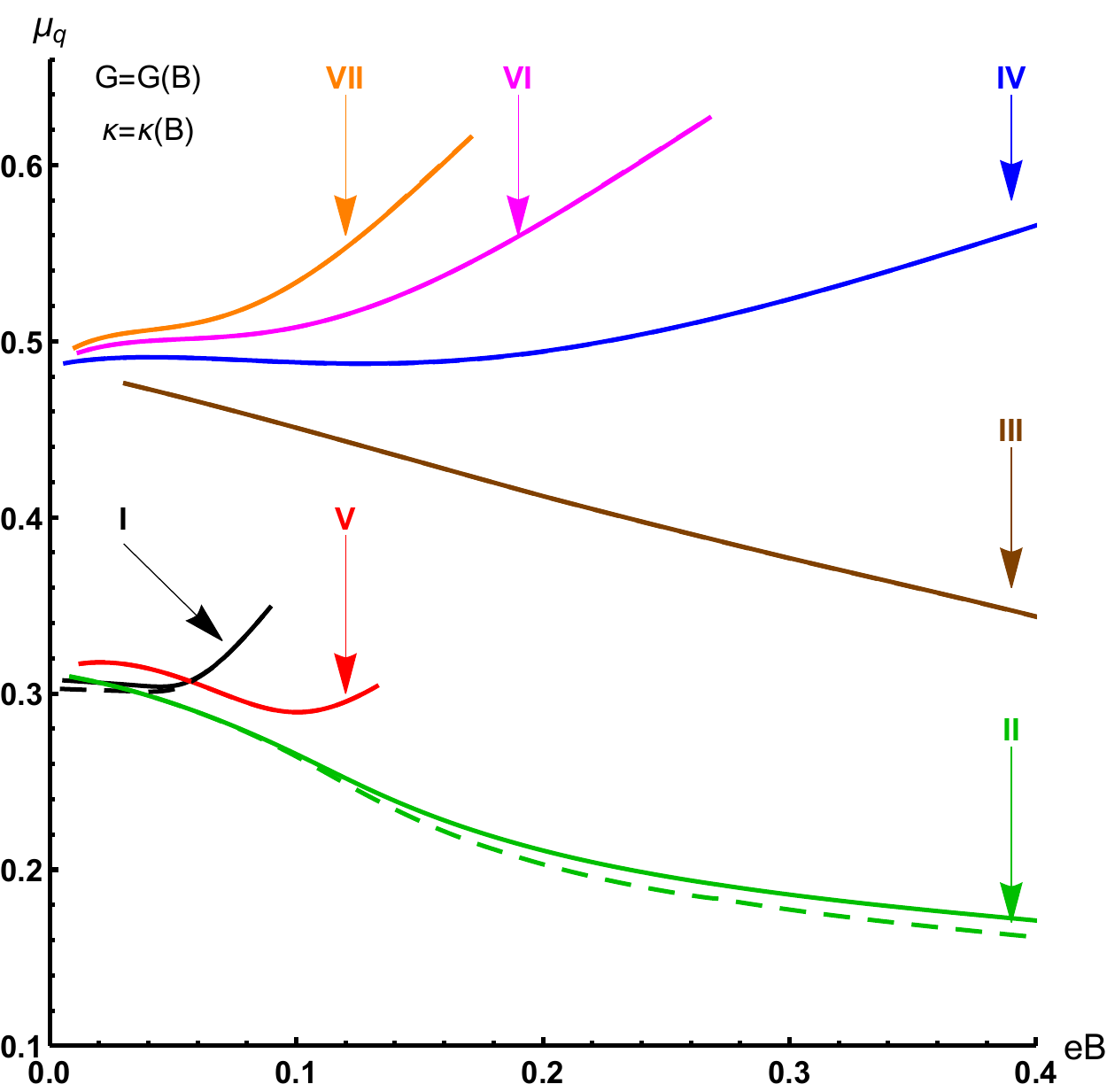}}
\caption{
On the left-hand side the external magnetic field strength ($[eB]=\mathrm{GeV}^2$) dependence of the temperature of the CEPs which are present in the $\mu-T$ phase diagram and on the right-hand side that of the corresponding quark chemical potentials ($[T]=[\mu_q]=\mathrm{GeV}$). Full lines correspond to the NJL case whereas dashed lines correspond to the extension to include the Polyakov loop. On the top row the coupling strengths for the model are kept fixed at their vacuum determined values whereas in the bottom row their magnetic field dependence is included.
}
\label{CEPsvseB}
\end{figure*}

In Fig. \ref{BDepEffectCEPPNJL} we group some of the results from Fig. \ref{CEPsvseB} for easier direct comparison of the results obtained with and without the magnetic field dependence on the couplings (we chose to use the results for the Polyakov loop extended version as the temperatures are higher). Due to the weaker spontaneous chiral symmetry breaking effects (lower NJL coupling) in the case where we consider magnetic field dependent couplings, both the temperature and the quark chemical potential of the CEPs are lower.

\begin{figure*}[!htb]
\center
\subfigure[]{\label{BDepEffectTCEPvsBPNJL}\includegraphics[width=0.49\textwidth]{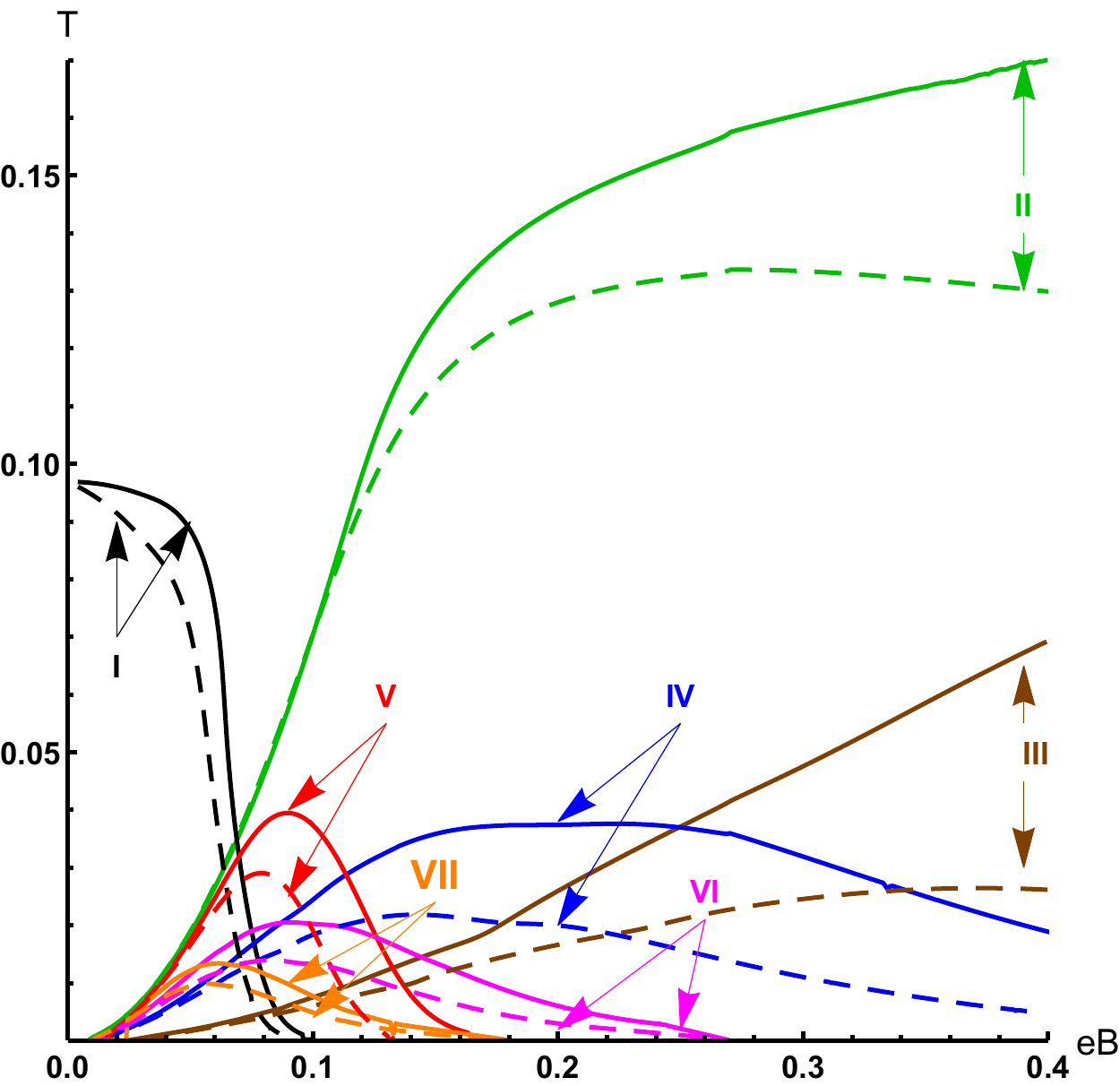}}
\subfigure[]{\label{BDepEffectmuCEPvsBPNJL}\includegraphics[width=0.49\textwidth]{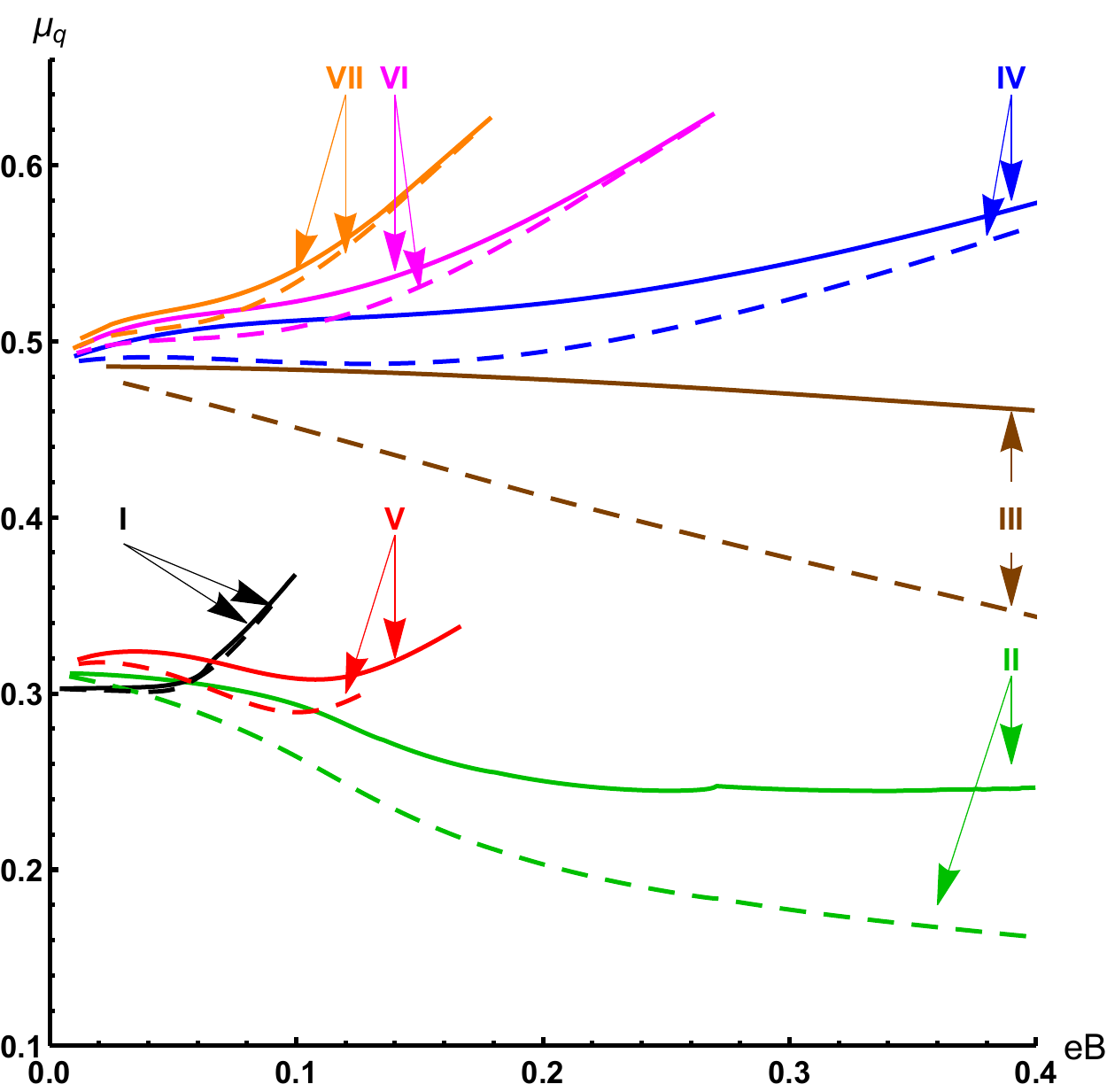}}
\caption{Comparison of the magnetic field dependence of the temperature (left-hand side panel) and quark chemical potential (right-hand side panel) of the CEP ($[T]=[\mu_q]=\mathrm{GeV}$ and $[eB]=\mathrm{GeV}^2$)in the cases with and without a magnetic field dependence on the interaction coupling strengths. Full lines correspond to the case where the couplings are kept fixed at the vacuum determined value whereas dashed lines correspond to the case where a magnetic field dependence in these couplings is considered.
}
\label{BDepEffectCEPPNJL}
\end{figure*}

In Figs. \ref{TmuCEPPPNJL} we combine the magnetic field dependence of the temperature and quark chemical potential  of the CEPs described above to show their path. We chose to depict that path for a range up to $eB=0.6~\mathrm{GeV}^2$. The main effects are the ones already described above and attributable to the weakening of the spontaneous chiral symmetry breaking (lower $G$): the temperatures reached by theses paths are lower in the magnetic field dependent couplings case (see for instance the height of the paths for CEPs $IV$, $VI$ and $VII$ and the dramatic lowering of the temperature of CEPs $II$ and $III$ which are the only ones to survive up to the superior limit of the studied range) and a shift towards lower chemical potentials occurs (deforming the first part of the CEP $V$ branch and opening the loop; notice also that the end of this loop is at a lower magnetic field strength in the case with magnetic field dependent couplings as can be seen in Fig. \ref{BDepEffectCEPPNJL}). 

Throughout all the preceding discussion one must bear in mind the existence of an underlying model validity limitation related to the regularization procedure and as such the consideration of arbitrarily high chemical potentials/magnetic field strengths/temperatures should be avoided.

\begin{figure*}[!htb]
\center
\subfigure[]{\label{TmuCEPPPNJLG0K0}\includegraphics[width=0.49\textwidth]{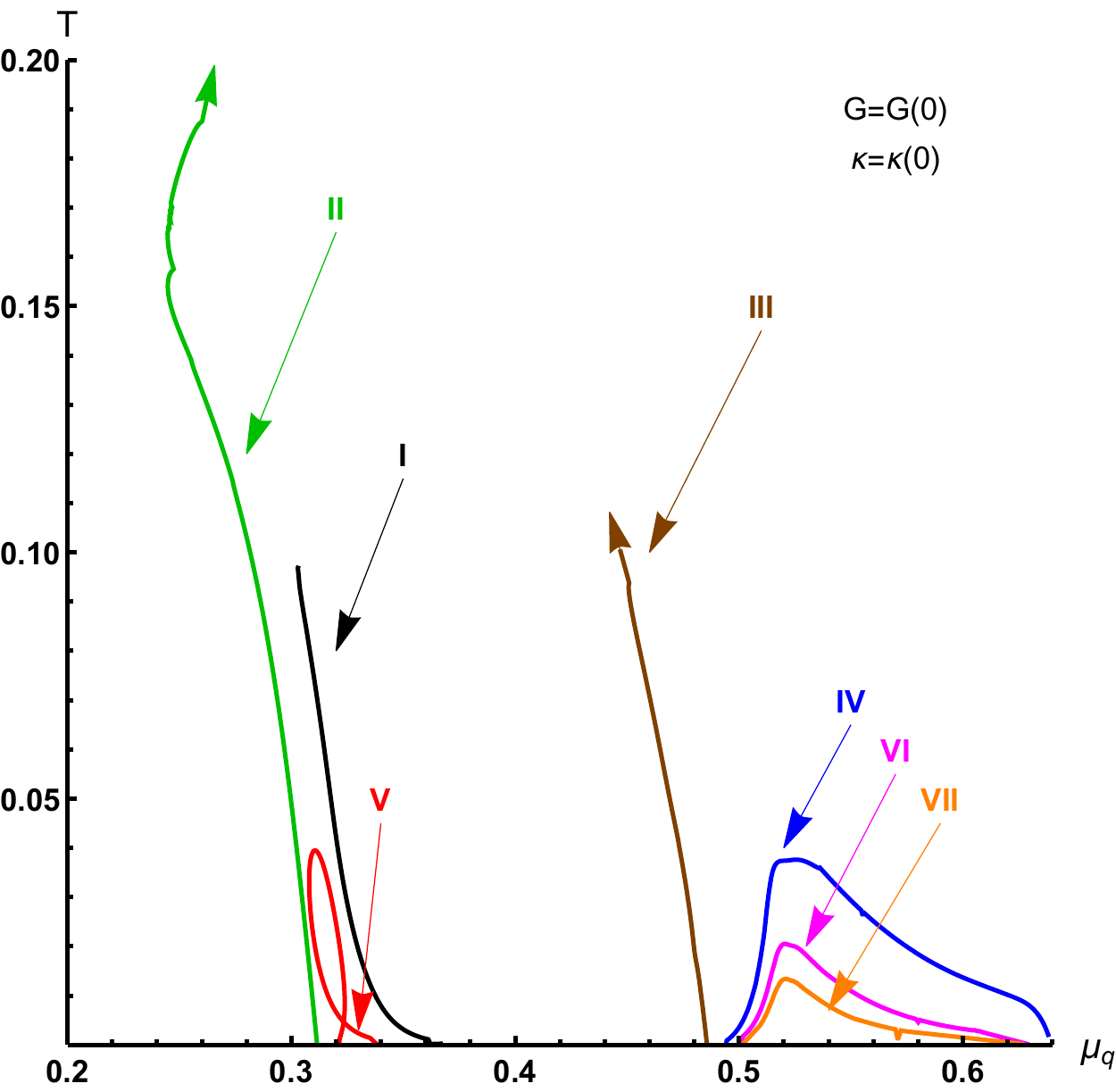}}
\subfigure[]{\label{TmuCEPPPNJLGBKB}\includegraphics[width=0.49\textwidth]{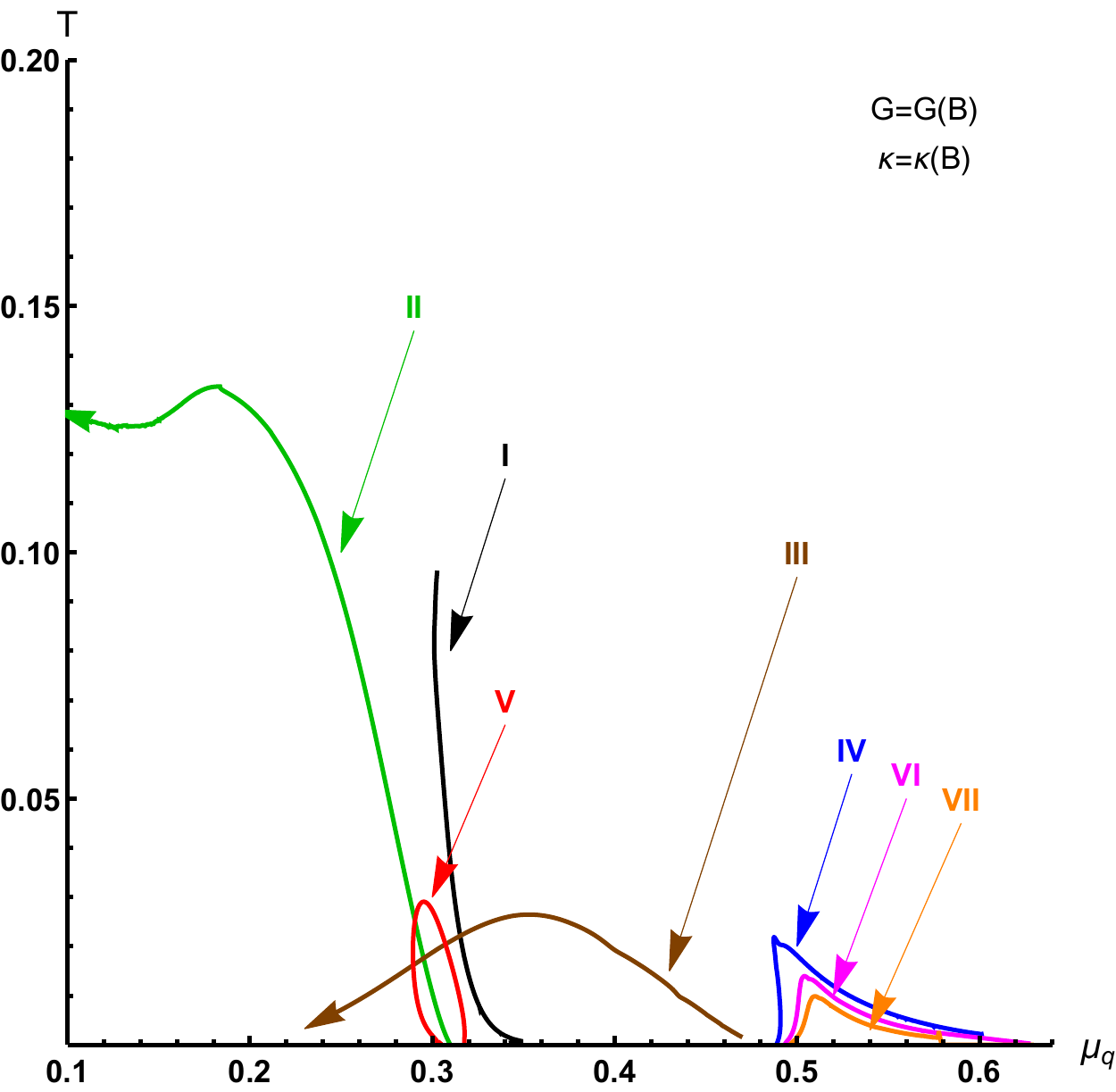}}
\caption{Paths described by the CEPs in the temperature-quark chemical potential plane ($[T]=[\mu_q]=\mathrm{GeV}$) for magnetic field strengths in the range $eB=0-0.6~\mathrm{GeV}^2$. Paths labeled $II$ and $III$ are open (the last point depicted is for $eB=0.6~\mathrm{GeV}^2$) as the line continues. On the left-hand side the case with coupling strengths fixed at vanishing magnetic field and on the right-hand side considering a magnetic field dependence.}
\label{TmuCEPPPNJL}
\end{figure*}

\section{Conclusions}\label{sec4}
\label{Conclusions}

In this work we have studied the implications for the phase diagram of strongly interacting matter, as described by the 't Hooft extended Nambu--Jona-Lasinio model, with and without the Polyakov Loop, of using the magnetic field dependence of the coupling strengths of the interactions derived previously \cite{Moreira:2020wau} from the magnetic field dependence of the dynamical quark masses on the \emph{up}, \emph{down} and \emph{strange} sectors (as reported in \cite{Endrodi:2019whh}). 
Our main findings were:
\begin{itemize}
\item  the weakening of the NJL interaction induced by the magnetic field results in shorter first-order phase transition lines as well as their shift towards lower chemical potentials being these effects expected as a direct result of the smaller dynamical chiral symmetry breaking effects.
\item the increase of relevance of the flavor mixing 't Hooft determinant (due to its increase with the magnetic field) results in a shift of the transition especially the one at the lowest chemical potential at vanishing temperature towards higher temperature and quark chemical potential. 
\item at a critical value of $eB\sim 0.01~\mathrm{GeV}^2$ six new CEPs were found to be introduced in the phase diagram. With further increase of the magnetic field strength, the original CEP plus several of the additional disappear with only two surviving in the studied range ($eB=0-0.6~\mathrm{GeV}^2$).
\item as expected, the main effect of the inclusion of the Polyakov loop in the phase diagram is that of raising the temperature of the transitions and therefore that of the CEP as well; in fact, as in the vanishing temperature limit it reverts to the case without Polyakov loop, the extension preserves the number of transitions and keeps the general features of the magnetic field dependence of the CEPs. It should be noted that in other implementations of this model, with a Polyakov loop dependence at vanishing temperature this might change \cite{CamaraPereira:2016oxs}. 
\end{itemize}

\section*{Acknowledgments}

This work was supported by a research grant under Project No. PTDC/FIS-NUC/29912/2017, funded by national funds through FCT (Fundação para a Ciência e a Tecnologia, I.P, Portugal) and co-financed by the European Regional Development Fund (ERDF) through the Portuguese Operational Program for Competitiveness and Internationalization, COMPETE 2020, by the project CERN/FIS-PAR/0040/2019 and by national funds from FCT, within the Projects No. UIDB\slash 04564\slash 2020 and No. UIDP\slash 04564\slash 2020. This study  was  financed  in  part  by Coordena\c c\~{a}o  de Aperfeiçoamento de Pessoal  de  N\'{\i}vel Superior-(CAPES-Brazil)-Finance  Code 001. T.E.R. thanks the support and hospitality of CFisUC and acknowledges Conselho Nacional de Desenvolvimento Cient\'{\i}fico e Tecnol\'{o}gico (CNPq-Brazil) and  Coordena\c c\~{a}o  de Aperfei\c coamento  de  Pessoal  de  N\'{\i}vel  Superior   (CAPES-Brazil) for   PhD grants at different periods of time.

\bibliographystyle{apsrev4-1}
\bibliography{PD_GKfit_lQCDMi}

\begin{thebibliography}{54}%
\makeatletter
\providecommand \@ifxundefined [1]{%
 \@ifx{#1\undefined}
}%
\providecommand \@ifnum [1]{%
 \ifnum #1\expandafter \@firstoftwo
 \else \expandafter \@secondoftwo
 \fi
}%
\providecommand \@ifx [1]{%
 \ifx #1\expandafter \@firstoftwo
 \else \expandafter \@secondoftwo
 \fi
}%
\providecommand \natexlab [1]{#1}%
\providecommand \enquote  [1]{``#1''}%
\providecommand \bibnamefont  [1]{#1}%
\providecommand \bibfnamefont [1]{#1}%
\providecommand \citenamefont [1]{#1}%
\providecommand \href@noop [0]{\@secondoftwo}%
\providecommand \href [0]{\begingroup \@sanitize@url \@href}%
\providecommand \@href[1]{\@@startlink{#1}\@@href}%
\providecommand \@@href[1]{\endgroup#1\@@endlink}%
\providecommand \@sanitize@url [0]{\catcode `\\12\catcode `\$12\catcode
  `\&12\catcode `\#12\catcode `\^12\catcode `\_12\catcode `\%12\relax}%
\providecommand \@@startlink[1]{}%
\providecommand \@@endlink[0]{}%
\providecommand \url  [0]{\begingroup\@sanitize@url \@url }%
\providecommand \@url [1]{\endgroup\@href {#1}{\urlprefix }}%
\providecommand \urlprefix  [0]{URL }%
\providecommand \Eprint [0]{\href }%
\providecommand \doibase [0]{http://dx.doi.org/}%
\providecommand \selectlanguage [0]{\@gobble}%
\providecommand \bibinfo  [0]{\@secondoftwo}%
\providecommand \bibfield  [0]{\@secondoftwo}%
\providecommand \translation [1]{[#1]}%
\providecommand \BibitemOpen [0]{}%
\providecommand \bibitemStop [0]{}%
\providecommand \bibitemNoStop [0]{.\EOS\space}%
\providecommand \EOS [0]{\spacefactor3000\relax}%
\providecommand \BibitemShut  [1]{\csname bibitem#1\endcsname}%
\let\auto@bib@innerbib\@empty
%</preamble>
\bibitem [{\citenamefont {Andronic}\ \emph {et~al.}(2018)\citenamefont
  {Andronic}, \citenamefont {Braun-Munzinger}, \citenamefont {Redlich},\ and\
  \citenamefont {Stachel}}]{Andronic:2017pug}%
  \BibitemOpen
  \bibfield  {author} {\bibinfo {author} {\bibfnamefont {A.}~\bibnamefont
  {Andronic}}, \bibinfo {author} {\bibfnamefont {P.}~\bibnamefont
  {Braun-Munzinger}}, \bibinfo {author} {\bibfnamefont {K.}~\bibnamefont
  {Redlich}}, \ and\ \bibinfo {author} {\bibfnamefont {J.}~\bibnamefont
  {Stachel}},\ }\href {\doibase 10.1038/s41586-018-0491-6} {\bibfield
  {journal} {\bibinfo  {journal} {Nature}\ }\textbf {\bibinfo {volume} {561}},\
  \bibinfo {pages} {321} (\bibinfo {year} {2018})},\ \Eprint
  {http://arxiv.org/abs/1710.09425} {arXiv:1710.09425 [nucl-th]} \BibitemShut
  {NoStop}%
\bibitem [{\citenamefont {Adamczyk}\ \emph {et~al.}(2017)\citenamefont
  {Adamczyk} \emph {et~al.}}]{Adamczyk:2017iwn}%
  \BibitemOpen
  \bibfield  {author} {\bibinfo {author} {\bibfnamefont {L.}~\bibnamefont
  {Adamczyk}} \emph {et~al.} (\bibinfo {collaboration} {STAR}),\ }\href
  {\doibase 10.1103/PhysRevC.96.044904} {\bibfield  {journal} {\bibinfo
  {journal} {Phys. Rev. C}\ }\textbf {\bibinfo {volume} {96}},\ \bibinfo
  {pages} {044904} (\bibinfo {year} {2017})},\ \Eprint
  {http://arxiv.org/abs/1701.07065} {arXiv:1701.07065 [nucl-ex]} \BibitemShut
  {NoStop}%
\bibitem [{\citenamefont {Yang}(2021)}]{Yang:2021lfe}%
  \BibitemOpen
  \bibfield  {author} {\bibinfo {author} {\bibfnamefont {Y.}~\bibnamefont
  {Yang}} (\bibinfo {collaboration} {STAR}),\ }\href {\doibase
  10.1016/j.nuclphysa.2020.121758} {\bibfield  {journal} {\bibinfo  {journal}
  {Nucl. Phys. A}\ }\textbf {\bibinfo {volume} {1005}},\ \bibinfo {pages}
  {121758} (\bibinfo {year} {2021})}\BibitemShut {NoStop}%
\bibitem [{\citenamefont {Yang}(2017)}]{Yang:2017llt}%
  \BibitemOpen
  \bibfield  {author} {\bibinfo {author} {\bibfnamefont {C.}~\bibnamefont
  {Yang}} (\bibinfo {collaboration} {STAR}),\ }\href {\doibase
  10.1016/j.nuclphysa.2017.05.042} {\bibfield  {journal} {\bibinfo  {journal}
  {Nucl. Phys. A}\ }\textbf {\bibinfo {volume} {967}},\ \bibinfo {pages} {800}
  (\bibinfo {year} {2017})}\BibitemShut {NoStop}%
\bibitem [{\citenamefont {Senger}(2017)}]{Senger:2017nvf}%
  \BibitemOpen
  \bibfield  {author} {\bibinfo {author} {\bibfnamefont {P.}~\bibnamefont
  {Senger}},\ }\href {\doibase 10.1088/1742-6596/798/1/012062} {\bibfield
  {journal} {\bibinfo  {journal} {J. Phys. Conf. Ser.}\ }\textbf {\bibinfo
  {volume} {798}},\ \bibinfo {pages} {012062} (\bibinfo {year}
  {2017})}\BibitemShut {NoStop}%
\bibitem [{\citenamefont {Senger}(2020)}]{Senger:2020fvj}%
  \BibitemOpen
  \bibfield  {author} {\bibinfo {author} {\bibfnamefont {P.}~\bibnamefont
  {Senger}},\ }\href {\doibase 10.7566/JPSCP.32.010092} {\bibfield  {journal}
  {\bibinfo  {journal} {JPS Conf. Proc.}\ }\textbf {\bibinfo {volume} {32}},\
  \bibinfo {pages} {010092} (\bibinfo {year} {2020})}\BibitemShut {NoStop}%
\bibitem [{\citenamefont {Kekelidze}\ \emph {et~al.}(2017)\citenamefont
  {Kekelidze}, \citenamefont {Kovalenko}, \citenamefont {Lednicky},
  \citenamefont {Matveev}, \citenamefont {Meshkov}, \citenamefont {Sorin},\
  and\ \citenamefont {Trubnikov}}]{Kekelidze:2017tgp}%
  \BibitemOpen
  \bibfield  {author} {\bibinfo {author} {\bibfnamefont {V.}~\bibnamefont
  {Kekelidze}}, \bibinfo {author} {\bibfnamefont {A.}~\bibnamefont
  {Kovalenko}}, \bibinfo {author} {\bibfnamefont {R.}~\bibnamefont {Lednicky}},
  \bibinfo {author} {\bibfnamefont {V.}~\bibnamefont {Matveev}}, \bibinfo
  {author} {\bibfnamefont {I.}~\bibnamefont {Meshkov}}, \bibinfo {author}
  {\bibfnamefont {A.}~\bibnamefont {Sorin}}, \ and\ \bibinfo {author}
  {\bibfnamefont {G.}~\bibnamefont {Trubnikov}},\ }\href {\doibase
  10.1016/j.nuclphysa.2017.06.031} {\bibfield  {journal} {\bibinfo  {journal}
  {Nucl. Phys. A}\ }\textbf {\bibinfo {volume} {967}},\ \bibinfo {pages} {884}
  (\bibinfo {year} {2017})}\BibitemShut {NoStop}%
\bibitem [{\citenamefont {Aoki}\ \emph {et~al.}(2006)\citenamefont {Aoki},
  \citenamefont {Endr\H{o}di}, \citenamefont {Fodor}, \citenamefont {Katz},\
  and\ \citenamefont {Szabo}}]{Aoki:2006we}%
  \BibitemOpen
  \bibfield  {author} {\bibinfo {author} {\bibfnamefont {Y.}~\bibnamefont
  {Aoki}}, \bibinfo {author} {\bibfnamefont {G.}~\bibnamefont {Endr\H{o}di}},
  \bibinfo {author} {\bibfnamefont {Z.}~\bibnamefont {Fodor}}, \bibinfo
  {author} {\bibfnamefont {S.~D.}\ \bibnamefont {Katz}}, \ and\ \bibinfo
  {author} {\bibfnamefont {K.~K.}\ \bibnamefont {Szabo}},\ }\href {\doibase
  10.1038/nature05120} {\bibfield  {journal} {\bibinfo  {journal} {Nature}\
  }\textbf {\bibinfo {volume} {443}},\ \bibinfo {pages} {675} (\bibinfo {year}
  {2006})},\ \Eprint {http://arxiv.org/abs/hep-lat/0611014}
  {arXiv:hep-lat/0611014} \BibitemShut {NoStop}%
%%CITATION = HEP-LAT/0611014;%%
\bibitem [{\citenamefont {Bazavov}\ \emph {et~al.}(2014)\citenamefont {Bazavov}
  \emph {et~al.}}]{Bazavov:2014pvz}%
  \BibitemOpen
  \bibfield  {author} {\bibinfo {author} {\bibfnamefont {A.}~\bibnamefont
  {Bazavov}} \emph {et~al.} (\bibinfo {collaboration} {HotQCD}),\ }\href
  {\doibase 10.1103/PhysRevD.90.094503} {\bibfield  {journal} {\bibinfo
  {journal} {Phys. Rev. D}\ }\textbf {\bibinfo {volume} {90}},\ \bibinfo
  {pages} {094503} (\bibinfo {year} {2014})},\ \Eprint
  {http://arxiv.org/abs/1407.6387} {arXiv:1407.6387 [hep-lat]} \BibitemShut
  {NoStop}%
\bibitem [{\citenamefont {Borsanyi}\ \emph {et~al.}(2010)\citenamefont
  {Borsanyi}, \citenamefont {Fodor}, \citenamefont {Hoelbling}, \citenamefont
  {Katz}, \citenamefont {Krieg}, \citenamefont {Ratti},\ and\ \citenamefont
  {Szabo}}]{Borsanyi:2010bp}%
  \BibitemOpen
  \bibfield  {author} {\bibinfo {author} {\bibfnamefont {S.}~\bibnamefont
  {Borsanyi}}, \bibinfo {author} {\bibfnamefont {Z.}~\bibnamefont {Fodor}},
  \bibinfo {author} {\bibfnamefont {C.}~\bibnamefont {Hoelbling}}, \bibinfo
  {author} {\bibfnamefont {S.~D.}\ \bibnamefont {Katz}}, \bibinfo {author}
  {\bibfnamefont {S.}~\bibnamefont {Krieg}}, \bibinfo {author} {\bibfnamefont
  {C.}~\bibnamefont {Ratti}}, \ and\ \bibinfo {author} {\bibfnamefont {K.~K.}\
  \bibnamefont {Szabo}} (\bibinfo {collaboration} {Wuppertal-Budapest}),\
  }\href {\doibase 10.1007/JHEP09(2010)073} {\bibfield  {journal} {\bibinfo
  {journal} {JHEP}\ }\textbf {\bibinfo {volume} {09}},\ \bibinfo {pages} {073}
  (\bibinfo {year} {2010})},\ \Eprint {http://arxiv.org/abs/1005.3508}
  {arXiv:1005.3508 [hep-lat]} \BibitemShut {NoStop}%
\bibitem [{\citenamefont {Borsanyi}\ \emph {et~al.}(2014)\citenamefont
  {Borsanyi}, \citenamefont {Fodor}, \citenamefont {Hoelbling}, \citenamefont
  {Katz}, \citenamefont {Krieg},\ and\ \citenamefont
  {Szabo}}]{Borsanyi:2013bia}%
  \BibitemOpen
  \bibfield  {author} {\bibinfo {author} {\bibfnamefont {S.}~\bibnamefont
  {Borsanyi}}, \bibinfo {author} {\bibfnamefont {Z.}~\bibnamefont {Fodor}},
  \bibinfo {author} {\bibfnamefont {C.}~\bibnamefont {Hoelbling}}, \bibinfo
  {author} {\bibfnamefont {S.~D.}\ \bibnamefont {Katz}}, \bibinfo {author}
  {\bibfnamefont {S.}~\bibnamefont {Krieg}}, \ and\ \bibinfo {author}
  {\bibfnamefont {K.~K.}\ \bibnamefont {Szabo}},\ }\href {\doibase
  10.1016/j.physletb.2014.01.007} {\bibfield  {journal} {\bibinfo  {journal}
  {Phys. Lett. B}\ }\textbf {\bibinfo {volume} {730}},\ \bibinfo {pages} {99}
  (\bibinfo {year} {2014})},\ \Eprint {http://arxiv.org/abs/1309.5258}
  {arXiv:1309.5258 [hep-lat]} \BibitemShut {NoStop}%
\bibitem [{\citenamefont {Stachel}\ \emph {et~al.}(2014)\citenamefont
  {Stachel}, \citenamefont {Andronic}, \citenamefont {Braun-Munzinger},\ and\
  \citenamefont {Redlich}}]{Stachel:2013zma}%
  \BibitemOpen
  \bibfield  {author} {\bibinfo {author} {\bibfnamefont {J.}~\bibnamefont
  {Stachel}}, \bibinfo {author} {\bibfnamefont {A.}~\bibnamefont {Andronic}},
  \bibinfo {author} {\bibfnamefont {P.}~\bibnamefont {Braun-Munzinger}}, \ and\
  \bibinfo {author} {\bibfnamefont {K.}~\bibnamefont {Redlich}},\ }\href
  {\doibase 10.1088/1742-6596/509/1/012019} {\bibfield  {journal} {\bibinfo
  {journal} {J. Phys. Conf. Ser.}\ }\textbf {\bibinfo {volume} {509}},\
  \bibinfo {pages} {012019} (\bibinfo {year} {2014})},\ \Eprint
  {http://arxiv.org/abs/1311.4662} {arXiv:1311.4662 [nucl-th]} \BibitemShut
  {NoStop}%
\bibitem [{\citenamefont {Dumitru}\ \emph {et~al.}(2005)\citenamefont
  {Dumitru}, \citenamefont {Pisarski},\ and\ \citenamefont
  {Zschiesche}}]{Dumitru:2005ng}%
  \BibitemOpen
  \bibfield  {author} {\bibinfo {author} {\bibfnamefont {A.}~\bibnamefont
  {Dumitru}}, \bibinfo {author} {\bibfnamefont {R.~D.}\ \bibnamefont
  {Pisarski}}, \ and\ \bibinfo {author} {\bibfnamefont {D.}~\bibnamefont
  {Zschiesche}},\ }\href {\doibase 10.1103/PhysRevD.72.065008} {\bibfield
  {journal} {\bibinfo  {journal} {Phys. Rev. D}\ }\textbf {\bibinfo {volume}
  {72}},\ \bibinfo {pages} {065008} (\bibinfo {year} {2005})},\ \Eprint
  {http://arxiv.org/abs/hep-ph/0505256} {arXiv:hep-ph/0505256} \BibitemShut
  {NoStop}%
\bibitem [{\citenamefont {Nambu}\ and\ \citenamefont
  {Jona-Lasinio}(1961{\natexlab{a}})}]{Nambu:1961fr}%
  \BibitemOpen
  \bibfield  {author} {\bibinfo {author} {\bibfnamefont {Y.}~\bibnamefont
  {Nambu}}\ and\ \bibinfo {author} {\bibfnamefont {G.}~\bibnamefont
  {Jona-Lasinio}},\ }\href {\doibase 10.1103/PhysRev.124.246} {\bibfield
  {journal} {\bibinfo  {journal} {Phys. Rev.}\ }\textbf {\bibinfo {volume}
  {124}},\ \bibinfo {pages} {246} (\bibinfo {year}
  {1961}{\natexlab{a}})}\BibitemShut {NoStop}%
%%CITATION = PHRVA,124,246;%%
\bibitem [{\citenamefont {Nambu}\ and\ \citenamefont
  {Jona-Lasinio}(1961{\natexlab{b}})}]{Nambu:1961tp}%
  \BibitemOpen
  \bibfield  {author} {\bibinfo {author} {\bibfnamefont {Y.}~\bibnamefont
  {Nambu}}\ and\ \bibinfo {author} {\bibfnamefont {G.}~\bibnamefont
  {Jona-Lasinio}},\ }\href {\doibase 10.1103/PhysRev.122.345} {\bibfield
  {journal} {\bibinfo  {journal} {Phys. Rev.}\ }\textbf {\bibinfo {volume}
  {122}},\ \bibinfo {pages} {345} (\bibinfo {year}
  {1961}{\natexlab{b}})}\BibitemShut {NoStop}%
%%CITATION = PHRVA,122,345;%%
\bibitem [{\citenamefont {Klevansky}(1992)}]{Klevansky:1992qe}%
  \BibitemOpen
  \bibfield  {author} {\bibinfo {author} {\bibfnamefont {S.~P.}\ \bibnamefont
  {Klevansky}},\ }\href {\doibase 10.1103/RevModPhys.64.649} {\bibfield
  {journal} {\bibinfo  {journal} {Rev. Mod. Phys.}\ }\textbf {\bibinfo {volume}
  {64}},\ \bibinfo {pages} {649} (\bibinfo {year} {1992})}\BibitemShut
  {NoStop}%
%%CITATION = RMPHA,64,649;%%
\bibitem [{\citenamefont {Hatsuda}\ and\ \citenamefont
  {Kunihiro}(1994)}]{Hatsuda:1994pi}%
  \BibitemOpen
  \bibfield  {author} {\bibinfo {author} {\bibfnamefont {T.}~\bibnamefont
  {Hatsuda}}\ and\ \bibinfo {author} {\bibfnamefont {T.}~\bibnamefont
  {Kunihiro}},\ }\href {\doibase 10.1016/0370-1573(94)90022-1} {\bibfield
  {journal} {\bibinfo  {journal} {Phys. Rept.}\ }\textbf {\bibinfo {volume}
  {247}},\ \bibinfo {pages} {221} (\bibinfo {year} {1994})},\ \Eprint
  {http://arxiv.org/abs/hep-ph/9401310} {arXiv:hep-ph/9401310} \BibitemShut
  {NoStop}%
%%CITATION = HEP-PH/9401310;%%
\bibitem [{\citenamefont {Fukushima}(2004)}]{Fukushima:2003fw}%
  \BibitemOpen
  \bibfield  {author} {\bibinfo {author} {\bibfnamefont {K.}~\bibnamefont
  {Fukushima}},\ }\href {\doibase 10.1016/j.physletb.2004.04.027} {\bibfield
  {journal} {\bibinfo  {journal} {Phys.Lett.}\ }\textbf {\bibinfo {volume}
  {B591}},\ \bibinfo {pages} {277} (\bibinfo {year} {2004})},\ \Eprint
  {http://arxiv.org/abs/hep-ph/0310121} {arXiv:hep-ph/0310121 [hep-ph]}
  \BibitemShut {NoStop}%
%%CITATION = HEP-PH/0310121;%%
\bibitem [{\citenamefont {Megias}\ \emph {et~al.}(2004)\citenamefont {Megias},
  \citenamefont {Ruiz~Arriola},\ and\ \citenamefont {Salcedo}}]{Megias:2003ui}%
  \BibitemOpen
  \bibfield  {author} {\bibinfo {author} {\bibfnamefont {E.}~\bibnamefont
  {Megias}}, \bibinfo {author} {\bibfnamefont {E.}~\bibnamefont
  {Ruiz~Arriola}}, \ and\ \bibinfo {author} {\bibfnamefont {L.~L.}\
  \bibnamefont {Salcedo}},\ }\href {\doibase 10.1103/PhysRevD.69.116003}
  {\bibfield  {journal} {\bibinfo  {journal} {Phys. Rev.}\ }\textbf {\bibinfo
  {volume} {D69}},\ \bibinfo {pages} {116003} (\bibinfo {year} {2004})},\
  \Eprint {http://arxiv.org/abs/hep-ph/0312133} {arXiv:hep-ph/0312133 [hep-ph]}
  \BibitemShut {NoStop}%
%%CITATION = HEP-PH/0312133;%%
\bibitem [{\citenamefont {Megias}\ \emph {et~al.}(2006)\citenamefont {Megias},
  \citenamefont {Ruiz~Arriola},\ and\ \citenamefont {Salcedo}}]{Megias:2004hj}%
  \BibitemOpen
  \bibfield  {author} {\bibinfo {author} {\bibfnamefont {E.}~\bibnamefont
  {Megias}}, \bibinfo {author} {\bibfnamefont {E.}~\bibnamefont
  {Ruiz~Arriola}}, \ and\ \bibinfo {author} {\bibfnamefont {L.}~\bibnamefont
  {Salcedo}},\ }\href {\doibase 10.1103/PhysRevD.74.065005} {\bibfield
  {journal} {\bibinfo  {journal} {Phys.Rev.}\ }\textbf {\bibinfo {volume}
  {D74}},\ \bibinfo {pages} {065005} (\bibinfo {year} {2006})},\ \Eprint
  {http://arxiv.org/abs/hep-ph/0412308} {arXiv:hep-ph/0412308 [hep-ph]}
  \BibitemShut {NoStop}%
%%CITATION = HEP-PH/0412308;%%
\bibitem [{\citenamefont {Roessner}\ \emph {et~al.}(2007)\citenamefont
  {Roessner}, \citenamefont {Ratti},\ and\ \citenamefont
  {Weise}}]{Roessner:2006xn}%
  \BibitemOpen
  \bibfield  {author} {\bibinfo {author} {\bibfnamefont {S.}~\bibnamefont
  {Roessner}}, \bibinfo {author} {\bibfnamefont {C.}~\bibnamefont {Ratti}}, \
  and\ \bibinfo {author} {\bibfnamefont {W.}~\bibnamefont {Weise}},\ }\href
  {\doibase 10.1103/PhysRevD.75.034007} {\bibfield  {journal} {\bibinfo
  {journal} {Phys. Rev.}\ }\textbf {\bibinfo {volume} {D75}},\ \bibinfo {pages}
  {034007} (\bibinfo {year} {2007})},\ \Eprint
  {http://arxiv.org/abs/hep-ph/0609281} {arXiv:hep-ph/0609281} \BibitemShut
  {NoStop}%
%%CITATION = HEP-PH/0609281;%%
\bibitem [{\citenamefont {C\^amara~Pereira}\ \emph {et~al.}(2020)\citenamefont
  {C\^amara~Pereira}, \citenamefont {Moreira},\ and\ \citenamefont
  {Costa}}]{CamaraPereira:2020rtu}%
  \BibitemOpen
  \bibfield  {author} {\bibinfo {author} {\bibfnamefont {R.}~\bibnamefont
  {C\^amara~Pereira}}, \bibinfo {author} {\bibfnamefont {J.}~\bibnamefont
  {Moreira}}, \ and\ \bibinfo {author} {\bibfnamefont {P.}~\bibnamefont
  {Costa}},\ }\href {\doibase 10.1140/epja/s10050-020-00223-8} {\bibfield
  {journal} {\bibinfo  {journal} {Eur. Phys. J. A}\ }\textbf {\bibinfo {volume}
  {56}},\ \bibinfo {pages} {214} (\bibinfo {year} {2020})},\ \Eprint
  {http://arxiv.org/abs/2006.02385} {arXiv:2006.02385 [hep-ph]} \BibitemShut
  {NoStop}%
\bibitem [{\citenamefont {Osipov}\ \emph {et~al.}(2006)\citenamefont {Osipov},
  \citenamefont {Hiller},\ and\ \citenamefont
  {da~Provid\^encia}}]{Osipov:2005tq}%
  \BibitemOpen
  \bibfield  {author} {\bibinfo {author} {\bibfnamefont {A.~A.}\ \bibnamefont
  {Osipov}}, \bibinfo {author} {\bibfnamefont {B.}~\bibnamefont {Hiller}}, \
  and\ \bibinfo {author} {\bibfnamefont {J.}~\bibnamefont {da~Provid\^encia}},\
  }\href {\doibase 10.1016/j.physletb.2006.01.008} {\bibfield  {journal}
  {\bibinfo  {journal} {Phys. Lett.}\ }\textbf {\bibinfo {volume} {B634}},\
  \bibinfo {pages} {48} (\bibinfo {year} {2006})},\ \Eprint
  {http://arxiv.org/abs/hep-ph/0508058} {arXiv:hep-ph/0508058} \BibitemShut
  {NoStop}%
%%CITATION = HEP-PH/0508058;%%
\bibitem [{\citenamefont {Osipov}\ \emph {et~al.}(2007)\citenamefont {Osipov},
  \citenamefont {Hiller}, \citenamefont {Blin},\ and\ \citenamefont
  {da~Provid\^encia}}]{Osipov:2006ns}%
  \BibitemOpen
  \bibfield  {author} {\bibinfo {author} {\bibfnamefont {A.~A.}\ \bibnamefont
  {Osipov}}, \bibinfo {author} {\bibfnamefont {B.}~\bibnamefont {Hiller}},
  \bibinfo {author} {\bibfnamefont {A.~H.}\ \bibnamefont {Blin}}, \ and\
  \bibinfo {author} {\bibfnamefont {J.}~\bibnamefont {da~Provid\^encia}},\
  }\href {\doibase 10.1016/j.aop.2006.08.004} {\bibfield  {journal} {\bibinfo
  {journal} {Annals Phys.}\ }\textbf {\bibinfo {volume} {322}},\ \bibinfo
  {pages} {2021} (\bibinfo {year} {2007})},\ \Eprint
  {http://arxiv.org/abs/hep-ph/0607066} {arXiv:hep-ph/0607066} \BibitemShut
  {NoStop}%
%%CITATION = HEP-PH/0607066;%%
\bibitem [{\citenamefont {Moreira}\ \emph {et~al.}(2015)\citenamefont
  {Moreira}, \citenamefont {Morais}, \citenamefont {Hiller}, \citenamefont
  {Osipov},\ and\ \citenamefont {Blin}}]{Moreira:2014qna}%
  \BibitemOpen
  \bibfield  {author} {\bibinfo {author} {\bibfnamefont {J.}~\bibnamefont
  {Moreira}}, \bibinfo {author} {\bibfnamefont {J.}~\bibnamefont {Morais}},
  \bibinfo {author} {\bibfnamefont {B.}~\bibnamefont {Hiller}}, \bibinfo
  {author} {\bibfnamefont {A.~A.}\ \bibnamefont {Osipov}}, \ and\ \bibinfo
  {author} {\bibfnamefont {A.~H.}\ \bibnamefont {Blin}},\ }\href {\doibase
  10.1103/PhysRevD.91.116003} {\bibfield  {journal} {\bibinfo  {journal} {Phys.
  Rev.}\ }\textbf {\bibinfo {volume} {D91}},\ \bibinfo {pages} {116003}
  (\bibinfo {year} {2015})},\ \Eprint {http://arxiv.org/abs/1409.0336}
  {arXiv:1409.0336 [hep-ph]} \BibitemShut {NoStop}%
%%CITATION = ARXIV:1409.0336;%%
\bibitem [{\citenamefont {Osipov}\ \emph
  {et~al.}(2013{\natexlab{a}})\citenamefont {Osipov}, \citenamefont {Hiller},\
  and\ \citenamefont {Blin}}]{Osipov:2012kk}%
  \BibitemOpen
  \bibfield  {author} {\bibinfo {author} {\bibfnamefont {A.}~\bibnamefont
  {Osipov}}, \bibinfo {author} {\bibfnamefont {B.}~\bibnamefont {Hiller}}, \
  and\ \bibinfo {author} {\bibfnamefont {A.}~\bibnamefont {Blin}},\ }\href
  {\doibase 10.1140/epja/i2013-13014-y} {\bibfield  {journal} {\bibinfo
  {journal} {Eur.Phys.J.}\ }\textbf {\bibinfo {volume} {A49}},\ \bibinfo
  {pages} {14} (\bibinfo {year} {2013}{\natexlab{a}})},\ \Eprint
  {http://arxiv.org/abs/1206.1920} {arXiv:1206.1920 [hep-ph]} \BibitemShut
  {NoStop}%
%%CITATION = ARXIV:1206.1920;%%
\bibitem [{\citenamefont {Osipov}\ \emph
  {et~al.}(2013{\natexlab{b}})\citenamefont {Osipov}, \citenamefont {Hiller},\
  and\ \citenamefont {Blin}}]{Osipov:2013fka}%
  \BibitemOpen
  \bibfield  {author} {\bibinfo {author} {\bibfnamefont {A.}~\bibnamefont
  {Osipov}}, \bibinfo {author} {\bibfnamefont {B.}~\bibnamefont {Hiller}}, \
  and\ \bibinfo {author} {\bibfnamefont {A.}~\bibnamefont {Blin}},\ }\href
  {\doibase 10.1103/PhysRevD.88.054032} {\bibfield  {journal} {\bibinfo
  {journal} {Phys.Rev.}\ }\textbf {\bibinfo {volume} {D88}},\ \bibinfo {pages}
  {054032} (\bibinfo {year} {2013}{\natexlab{b}})},\ \Eprint
  {http://arxiv.org/abs/1309.2497} {arXiv:1309.2497 [hep-ph]} \BibitemShut
  {NoStop}%
%%CITATION = ARXIV:1309.2497;%%
\bibitem [{\citenamefont {Skokov}\ \emph {et~al.}(2009)\citenamefont {Skokov},
  \citenamefont {Illarionov},\ and\ \citenamefont {Toneev}}]{Skokov:2009qp}%
  \BibitemOpen
  \bibfield  {author} {\bibinfo {author} {\bibfnamefont {V.}~\bibnamefont
  {Skokov}}, \bibinfo {author} {\bibfnamefont {A.~{\relax Yu}.}\ \bibnamefont
  {Illarionov}}, \ and\ \bibinfo {author} {\bibfnamefont {V.}~\bibnamefont
  {Toneev}},\ }\href {\doibase 10.1142/S0217751X09047570} {\bibfield  {journal}
  {\bibinfo  {journal} {Int. J. Mod. Phys.}\ }\textbf {\bibinfo {volume}
  {A24}},\ \bibinfo {pages} {5925} (\bibinfo {year} {2009})},\ \Eprint
  {http://arxiv.org/abs/0907.1396} {arXiv:0907.1396 [nucl-th]} \BibitemShut
  {NoStop}%
%%CITATION = ARXIV:0907.1396;%%
\bibitem [{\citenamefont {Menezes}\ \emph
  {et~al.}(2009{\natexlab{a}})\citenamefont {Menezes}, \citenamefont
  {Benghi~Pinto}, \citenamefont {Avancini}, \citenamefont {Perez~Martinez},\
  and\ \citenamefont {Provid\^encia}}]{Menezes:2008qt}%
  \BibitemOpen
  \bibfield  {author} {\bibinfo {author} {\bibfnamefont {D.~P.}\ \bibnamefont
  {Menezes}}, \bibinfo {author} {\bibfnamefont {M.}~\bibnamefont
  {Benghi~Pinto}}, \bibinfo {author} {\bibfnamefont {S.~S.}\ \bibnamefont
  {Avancini}}, \bibinfo {author} {\bibfnamefont {A.}~\bibnamefont
  {Perez~Martinez}}, \ and\ \bibinfo {author} {\bibfnamefont {C.}~\bibnamefont
  {Provid\^encia}},\ }\href {\doibase 10.1103/PhysRevC.79.035807} {\bibfield
  {journal} {\bibinfo  {journal} {Phys. Rev.}\ }\textbf {\bibinfo {volume}
  {C79}},\ \bibinfo {pages} {035807} (\bibinfo {year} {2009}{\natexlab{a}})},\
  \Eprint {http://arxiv.org/abs/0811.3361} {arXiv:0811.3361 [nucl-th]}
  \BibitemShut {NoStop}%
%%CITATION = ARXIV:0811.3361;%%
\bibitem [{\citenamefont {Costa}\ \emph {et~al.}(2015)\citenamefont {Costa},
  \citenamefont {Ferreira}, \citenamefont {Menezes}, \citenamefont {Moreira},\
  and\ \citenamefont {Provid\^encia}}]{Costa:2015bza}%
  \BibitemOpen
  \bibfield  {author} {\bibinfo {author} {\bibfnamefont {P.}~\bibnamefont
  {Costa}}, \bibinfo {author} {\bibfnamefont {M.}~\bibnamefont {Ferreira}},
  \bibinfo {author} {\bibfnamefont {D.~P.}\ \bibnamefont {Menezes}}, \bibinfo
  {author} {\bibfnamefont {J.}~\bibnamefont {Moreira}}, \ and\ \bibinfo
  {author} {\bibfnamefont {C.}~\bibnamefont {Provid\^encia}},\ }\href {\doibase
  10.1103/PhysRevD.92.036012} {\bibfield  {journal} {\bibinfo  {journal} {Phys.
  Rev. D}\ }\textbf {\bibinfo {volume} {92}},\ \bibinfo {pages} {036012}
  (\bibinfo {year} {2015})},\ \Eprint {http://arxiv.org/abs/1508.07870}
  {arXiv:1508.07870 [hep-ph]} \BibitemShut {NoStop}%
\bibitem [{\citenamefont {Costa}(2016)}]{Costa:2016vbb}%
  \BibitemOpen
  \bibfield  {author} {\bibinfo {author} {\bibfnamefont {P.}~\bibnamefont
  {Costa}},\ }\href {\doibase 10.1103/PhysRevD.93.114035} {\bibfield  {journal}
  {\bibinfo  {journal} {Phys. Rev. D}\ }\textbf {\bibinfo {volume} {93}},\
  \bibinfo {pages} {114035} (\bibinfo {year} {2016})},\ \Eprint
  {http://arxiv.org/abs/1610.06433} {arXiv:1610.06433 [nucl-th]} \BibitemShut
  {NoStop}%
\bibitem [{\citenamefont {Costa}\ \emph {et~al.}(2014)\citenamefont {Costa},
  \citenamefont {Ferreira}, \citenamefont {Hansen}, \citenamefont {Menezes},\
  and\ \citenamefont {Provid\^encia}}]{Costa:2013zca}%
  \BibitemOpen
  \bibfield  {author} {\bibinfo {author} {\bibfnamefont {P.}~\bibnamefont
  {Costa}}, \bibinfo {author} {\bibfnamefont {M.}~\bibnamefont {Ferreira}},
  \bibinfo {author} {\bibfnamefont {H.}~\bibnamefont {Hansen}}, \bibinfo
  {author} {\bibfnamefont {D.~P.}\ \bibnamefont {Menezes}}, \ and\ \bibinfo
  {author} {\bibfnamefont {C.}~\bibnamefont {Provid\^encia}},\ }\href {\doibase
  10.1103/PhysRevD.89.056013} {\bibfield  {journal} {\bibinfo  {journal} {Phys.
  Rev. D}\ }\textbf {\bibinfo {volume} {89}},\ \bibinfo {pages} {056013}
  (\bibinfo {year} {2014})},\ \Eprint {http://arxiv.org/abs/1307.7894}
  {arXiv:1307.7894 [hep-ph]} \BibitemShut {NoStop}%
\bibitem [{\citenamefont {Ferreira}\ \emph
  {et~al.}(2018{\natexlab{a}})\citenamefont {Ferreira}, \citenamefont {Costa},\
  and\ \citenamefont {Provid\^encia}}]{Ferreira:2017wtx}%
  \BibitemOpen
  \bibfield  {author} {\bibinfo {author} {\bibfnamefont {M.}~\bibnamefont
  {Ferreira}}, \bibinfo {author} {\bibfnamefont {P.}~\bibnamefont {Costa}}, \
  and\ \bibinfo {author} {\bibfnamefont {C.}~\bibnamefont {Provid\^encia}},\
  }\href {\doibase 10.1103/PhysRevD.97.014014} {\bibfield  {journal} {\bibinfo
  {journal} {Phys. Rev. D}\ }\textbf {\bibinfo {volume} {97}},\ \bibinfo
  {pages} {014014} (\bibinfo {year} {2018}{\natexlab{a}})},\ \Eprint
  {http://arxiv.org/abs/1712.08378} {arXiv:1712.08378 [hep-ph]} \BibitemShut
  {NoStop}%
\bibitem [{\citenamefont {Ferreira}\ \emph
  {et~al.}(2018{\natexlab{b}})\citenamefont {Ferreira}, \citenamefont {Costa},\
  and\ \citenamefont {Provid\^encia}}]{Ferreira:2018pux}%
  \BibitemOpen
  \bibfield  {author} {\bibinfo {author} {\bibfnamefont {M.}~\bibnamefont
  {Ferreira}}, \bibinfo {author} {\bibfnamefont {P.}~\bibnamefont {Costa}}, \
  and\ \bibinfo {author} {\bibfnamefont {C.}~\bibnamefont {Provid\^encia}},\
  }\href {\doibase 10.1103/PhysRevD.98.034003} {\bibfield  {journal} {\bibinfo
  {journal} {Phys. Rev. D}\ }\textbf {\bibinfo {volume} {98}},\ \bibinfo
  {pages} {034003} (\bibinfo {year} {2018}{\natexlab{b}})},\ \Eprint
  {http://arxiv.org/abs/1806.05758} {arXiv:1806.05758 [hep-ph]} \BibitemShut
  {NoStop}%
\bibitem [{\citenamefont {Ebert}\ \emph {et~al.}(2000)\citenamefont {Ebert},
  \citenamefont {Klimenko}, \citenamefont {Vdovichenko},\ and\ \citenamefont
  {Vshivtsev}}]{Ebert:1999ht}%
  \BibitemOpen
  \bibfield  {author} {\bibinfo {author} {\bibfnamefont {D.}~\bibnamefont
  {Ebert}}, \bibinfo {author} {\bibfnamefont {K.~G.}\ \bibnamefont {Klimenko}},
  \bibinfo {author} {\bibfnamefont {M.~A.}\ \bibnamefont {Vdovichenko}}, \ and\
  \bibinfo {author} {\bibfnamefont {A.~S.}\ \bibnamefont {Vshivtsev}},\ }\href
  {\doibase 10.1103/PhysRevD.61.025005} {\bibfield  {journal} {\bibinfo
  {journal} {Phys. Rev.}\ }\textbf {\bibinfo {volume} {D61}},\ \bibinfo {pages}
  {025005} (\bibinfo {year} {2000})},\ \Eprint
  {http://arxiv.org/abs/hep-ph/9905253} {arXiv:hep-ph/9905253 [hep-ph]}
  \BibitemShut {NoStop}%
%%CITATION = HEP-PH/9905253;%%
\bibitem [{\citenamefont {Denke}\ and\ \citenamefont
  {Pinto}(2013)}]{Denke:2013gha}%
  \BibitemOpen
  \bibfield  {author} {\bibinfo {author} {\bibfnamefont {R.~Z.}\ \bibnamefont
  {Denke}}\ and\ \bibinfo {author} {\bibfnamefont {M.~B.}\ \bibnamefont
  {Pinto}},\ }\href {\doibase 10.1103/PhysRevD.88.056008} {\bibfield  {journal}
  {\bibinfo  {journal} {Phys. Rev. D}\ }\textbf {\bibinfo {volume} {88}},\
  \bibinfo {pages} {056008} (\bibinfo {year} {2013})},\ \Eprint
  {http://arxiv.org/abs/1306.6246} {arXiv:1306.6246 [hep-ph]} \BibitemShut
  {NoStop}%
\bibitem [{\citenamefont {Bruckmann}\ \emph {et~al.}(2013)\citenamefont
  {Bruckmann}, \citenamefont {Endr{\H{o}}di},\ and\ \citenamefont
  {Kovacs}}]{Bruckmann:2013oba}%
  \BibitemOpen
  \bibfield  {author} {\bibinfo {author} {\bibfnamefont {F.}~\bibnamefont
  {Bruckmann}}, \bibinfo {author} {\bibfnamefont {G.}~\bibnamefont
  {Endr{\H{o}}di}}, \ and\ \bibinfo {author} {\bibfnamefont {T.~G.}\
  \bibnamefont {Kovacs}},\ }\href {\doibase 10.1007/JHEP04(2013)112} {\bibfield
   {journal} {\bibinfo  {journal} {JHEP}\ }\textbf {\bibinfo {volume} {04}},\
  \bibinfo {pages} {112} (\bibinfo {year} {2013})},\ \Eprint
  {http://arxiv.org/abs/1303.3972} {arXiv:1303.3972 [hep-lat]} \BibitemShut
  {NoStop}%
%%CITATION = ARXIV:1303.3972;%%
\bibitem [{\citenamefont {Bali}\ \emph
  {et~al.}(2012{\natexlab{a}})\citenamefont {Bali}, \citenamefont {Bruckmann},
  \citenamefont {Endr\H{o}di}, \citenamefont {Fodor}, \citenamefont {Katz},
  \citenamefont {Krieg}, \citenamefont {Sch\"afer},\ and\ \citenamefont
  {Szabo}}]{Bali:2011qj}%
  \BibitemOpen
  \bibfield  {author} {\bibinfo {author} {\bibfnamefont {G.~S.}\ \bibnamefont
  {Bali}}, \bibinfo {author} {\bibfnamefont {F.}~\bibnamefont {Bruckmann}},
  \bibinfo {author} {\bibfnamefont {G.}~\bibnamefont {Endr\H{o}di}}, \bibinfo
  {author} {\bibfnamefont {Z.}~\bibnamefont {Fodor}}, \bibinfo {author}
  {\bibfnamefont {S.~D.}\ \bibnamefont {Katz}}, \bibinfo {author}
  {\bibfnamefont {S.}~\bibnamefont {Krieg}}, \bibinfo {author} {\bibfnamefont
  {A.}~\bibnamefont {Sch\"afer}}, \ and\ \bibinfo {author} {\bibfnamefont
  {K.~K.}\ \bibnamefont {Szabo}},\ }\href {\doibase 10.1007/JHEP02(2012)044}
  {\bibfield  {journal} {\bibinfo  {journal} {JHEP}\ }\textbf {\bibinfo
  {volume} {02}},\ \bibinfo {pages} {044} (\bibinfo {year}
  {2012}{\natexlab{a}})},\ \Eprint {http://arxiv.org/abs/1111.4956}
  {arXiv:1111.4956 [hep-lat]} \BibitemShut {NoStop}%
%%CITATION = ARXIV:1111.4956;%%
\bibitem [{\citenamefont {Bali}\ \emph
  {et~al.}(2012{\natexlab{b}})\citenamefont {Bali}, \citenamefont {Bruckmann},
  \citenamefont {Endr\H{o}di}, \citenamefont {Fodor}, \citenamefont {Katz},\
  and\ \citenamefont {Sch\"afer}}]{Bali:2012zg}%
  \BibitemOpen
  \bibfield  {author} {\bibinfo {author} {\bibfnamefont {G.~S.}\ \bibnamefont
  {Bali}}, \bibinfo {author} {\bibfnamefont {F.}~\bibnamefont {Bruckmann}},
  \bibinfo {author} {\bibfnamefont {G.}~\bibnamefont {Endr\H{o}di}}, \bibinfo
  {author} {\bibfnamefont {Z.}~\bibnamefont {Fodor}}, \bibinfo {author}
  {\bibfnamefont {S.~D.}\ \bibnamefont {Katz}}, \ and\ \bibinfo {author}
  {\bibfnamefont {A.}~\bibnamefont {Sch\"afer}},\ }\href {\doibase
  10.1103/PhysRevD.86.071502} {\bibfield  {journal} {\bibinfo  {journal} {Phys.
  Rev.}\ }\textbf {\bibinfo {volume} {D86}},\ \bibinfo {pages} {071502}
  (\bibinfo {year} {2012}{\natexlab{b}})},\ \Eprint
  {http://arxiv.org/abs/1206.4205} {arXiv:1206.4205 [hep-lat]} \BibitemShut
  {NoStop}%
%%CITATION = ARXIV:1206.4205;%%
\bibitem [{\citenamefont {Endr\H{o}di}(2015)}]{Endrodi:2015oba}%
  \BibitemOpen
  \bibfield  {author} {\bibinfo {author} {\bibfnamefont {G.}~\bibnamefont
  {Endr\H{o}di}},\ }\href {\doibase 10.1007/JHEP07(2015)173} {\bibfield
  {journal} {\bibinfo  {journal} {JHEP}\ }\textbf {\bibinfo {volume} {07}},\
  \bibinfo {pages} {173} (\bibinfo {year} {2015})},\ \Eprint
  {http://arxiv.org/abs/1504.08280} {arXiv:1504.08280 [hep-lat]} \BibitemShut
  {NoStop}%
%%CITATION = ARXIV:1504.08280;%%
\bibitem [{\citenamefont {Endr\H{o}di}\ and\ \citenamefont
  {Markó}(2019)}]{Endrodi:2019whh}%
  \BibitemOpen
  \bibfield  {author} {\bibinfo {author} {\bibfnamefont {G.}~\bibnamefont
  {Endr\H{o}di}}\ and\ \bibinfo {author} {\bibfnamefont {G.}~\bibnamefont
  {Markó}},\ }\href {\doibase 10.1007/JHEP08(2019)036} {\bibfield  {journal}
  {\bibinfo  {journal} {JHEP}\ }\textbf {\bibinfo {volume} {08}},\ \bibinfo
  {pages} {036} (\bibinfo {year} {2019})},\ \Eprint
  {http://arxiv.org/abs/1905.02103} {arXiv:1905.02103 [hep-lat]} \BibitemShut
  {NoStop}%
%%CITATION = ARXIV:1905.02103;%%
\bibitem [{\citenamefont {Moreira}\ \emph {et~al.}(2020)\citenamefont
  {Moreira}, \citenamefont {Costa},\ and\ \citenamefont
  {Restrepo}}]{Moreira:2020wau}%
  \BibitemOpen
  \bibfield  {author} {\bibinfo {author} {\bibfnamefont {J.}~\bibnamefont
  {Moreira}}, \bibinfo {author} {\bibfnamefont {P.}~\bibnamefont {Costa}}, \
  and\ \bibinfo {author} {\bibfnamefont {T.~E.}\ \bibnamefont {Restrepo}},\
  }\href {\doibase 10.1103/PhysRevD.102.014032} {\bibfield  {journal} {\bibinfo
   {journal} {Phys. Rev. D}\ }\textbf {\bibinfo {volume} {102}},\ \bibinfo
  {pages} {014032} (\bibinfo {year} {2020})},\ \Eprint
  {http://arxiv.org/abs/2005.07049} {arXiv:2005.07049 [hep-ph]} \BibitemShut
  {NoStop}%
\bibitem [{\citenamefont {Ferreira}\ \emph
  {et~al.}(2014{\natexlab{a}})\citenamefont {Ferreira}, \citenamefont {Costa},
  \citenamefont {Menezes}, \citenamefont {Providência},\ and\ \citenamefont
  {Scoccola}}]{Ferreira:2013tba}%
  \BibitemOpen
  \bibfield  {author} {\bibinfo {author} {\bibfnamefont {M.}~\bibnamefont
  {Ferreira}}, \bibinfo {author} {\bibfnamefont {P.}~\bibnamefont {Costa}},
  \bibinfo {author} {\bibfnamefont {D.~P.}\ \bibnamefont {Menezes}}, \bibinfo
  {author} {\bibfnamefont {C.}~\bibnamefont {Providência}}, \ and\ \bibinfo
  {author} {\bibfnamefont {N.}~\bibnamefont {Scoccola}},\ }\href {\doibase
  10.1103/PhysRevD.89.016002, 10.1103/PhysRevD.89.019902} {\bibfield  {journal}
  {\bibinfo  {journal} {Phys. Rev.}\ }\textbf {\bibinfo {volume} {D89}},\
  \bibinfo {pages} {016002} (\bibinfo {year} {2014}{\natexlab{a}})},\ \bibinfo
  {note} {[Addendum: Phys. Rev.D89,no.1,019902(2014)]},\ \Eprint
  {http://arxiv.org/abs/1305.4751} {arXiv:1305.4751 [hep-ph]} \BibitemShut
  {NoStop}%
%%CITATION = ARXIV:1305.4751;%%
\bibitem [{\citenamefont {Ferreira}\ \emph
  {et~al.}(2014{\natexlab{b}})\citenamefont {Ferreira}, \citenamefont {Costa},
  \citenamefont {Lourenço}, \citenamefont {Frederico},\ and\ \citenamefont
  {Providência}}]{Ferreira:2014kpa}%
  \BibitemOpen
  \bibfield  {author} {\bibinfo {author} {\bibfnamefont {M.}~\bibnamefont
  {Ferreira}}, \bibinfo {author} {\bibfnamefont {P.}~\bibnamefont {Costa}},
  \bibinfo {author} {\bibfnamefont {O.}~\bibnamefont {Lourenço}}, \bibinfo
  {author} {\bibfnamefont {T.}~\bibnamefont {Frederico}}, \ and\ \bibinfo
  {author} {\bibfnamefont {C.}~\bibnamefont {Providência}},\ }\href {\doibase
  10.1103/PhysRevD.89.116011} {\bibfield  {journal} {\bibinfo  {journal}
  {Phys.\ Rev.\ D}\ }\textbf {\bibinfo {volume} {89}},\ \bibinfo {pages}
  {116011} (\bibinfo {year} {2014}{\natexlab{b}})},\ \Eprint
  {http://arxiv.org/abs/1404.5577} {arXiv:1404.5577 [hep-ph]} \BibitemShut
  {NoStop}%
\bibitem [{\citenamefont {Farias}\ \emph {et~al.}(2017)\citenamefont {Farias},
  \citenamefont {Timoteo}, \citenamefont {Avancini}, \citenamefont {Pinto},\
  and\ \citenamefont {Krein}}]{Farias:2016gmy}%
  \BibitemOpen
  \bibfield  {author} {\bibinfo {author} {\bibfnamefont {R.}~\bibnamefont
  {Farias}}, \bibinfo {author} {\bibfnamefont {V.}~\bibnamefont {Timoteo}},
  \bibinfo {author} {\bibfnamefont {S.}~\bibnamefont {Avancini}}, \bibinfo
  {author} {\bibfnamefont {M.}~\bibnamefont {Pinto}}, \ and\ \bibinfo {author}
  {\bibfnamefont {G.}~\bibnamefont {Krein}},\ }\href {\doibase
  10.1140/epja/i2017-12320-8} {\bibfield  {journal} {\bibinfo  {journal} {Eur.
  Phys. J. A}\ }\textbf {\bibinfo {volume} {53}},\ \bibinfo {pages} {101}
  (\bibinfo {year} {2017})},\ \Eprint {http://arxiv.org/abs/1603.03847}
  {arXiv:1603.03847 [hep-ph]} \BibitemShut {NoStop}%
\bibitem [{\citenamefont {Ratti}\ \emph {et~al.}(2006)\citenamefont {Ratti},
  \citenamefont {Thaler},\ and\ \citenamefont {Weise}}]{Ratti:2005jh}%
  \BibitemOpen
  \bibfield  {author} {\bibinfo {author} {\bibfnamefont {C.}~\bibnamefont
  {Ratti}}, \bibinfo {author} {\bibfnamefont {M.~A.}\ \bibnamefont {Thaler}}, \
  and\ \bibinfo {author} {\bibfnamefont {W.}~\bibnamefont {Weise}},\ }\href
  {\doibase 10.1103/PhysRevD.73.014019} {\bibfield  {journal} {\bibinfo
  {journal} {Phys. Rev.}\ }\textbf {\bibinfo {volume} {D73}},\ \bibinfo {pages}
  {014019} (\bibinfo {year} {2006})},\ \Eprint
  {http://arxiv.org/abs/hep-ph/0506234} {arXiv:hep-ph/0506234} \BibitemShut
  {NoStop}%
%%CITATION = HEP-PH/0506234;%%
\bibitem [{\citenamefont {Ratti}\ \emph {et~al.}(2007)\citenamefont {Ratti},
  \citenamefont {Roessner}, \citenamefont {Thaler},\ and\ \citenamefont
  {Weise}}]{Ratti:2006wg}%
  \BibitemOpen
  \bibfield  {author} {\bibinfo {author} {\bibfnamefont {C.}~\bibnamefont
  {Ratti}}, \bibinfo {author} {\bibfnamefont {S.}~\bibnamefont {Roessner}},
  \bibinfo {author} {\bibfnamefont {M.~A.}\ \bibnamefont {Thaler}}, \ and\
  \bibinfo {author} {\bibfnamefont {W.}~\bibnamefont {Weise}},\ }\bibfield
  {booktitle} {\emph {\bibinfo {booktitle} {{Proceedings, Workshop for Young
  Scientists on the Physics of Ultrarelativistic Nucleus-Nucleus Collisions
  (Hot Quarks 2006): Villasimius, Italy, May 15-20, 2006}}},\ }\href {\doibase
  10.1140/epjc/s10052-006-0065-x} {\bibfield  {journal} {\bibinfo  {journal}
  {Eur. Phys. J.}\ }\textbf {\bibinfo {volume} {C49}},\ \bibinfo {pages} {213}
  (\bibinfo {year} {2007})},\ \Eprint {http://arxiv.org/abs/hep-ph/0609218}
  {arXiv:hep-ph/0609218 [hep-ph]} \BibitemShut {NoStop}%
%%CITATION = HEP-PH/0609218;%%
\bibitem [{\citenamefont {Kaczmarek}\ \emph {et~al.}(2002)\citenamefont
  {Kaczmarek}, \citenamefont {Karsch}, \citenamefont {Petreczky},\ and\
  \citenamefont {Zantow}}]{Kaczmarek:2002mc}%
  \BibitemOpen
  \bibfield  {author} {\bibinfo {author} {\bibfnamefont {O.}~\bibnamefont
  {Kaczmarek}}, \bibinfo {author} {\bibfnamefont {F.}~\bibnamefont {Karsch}},
  \bibinfo {author} {\bibfnamefont {P.}~\bibnamefont {Petreczky}}, \ and\
  \bibinfo {author} {\bibfnamefont {F.}~\bibnamefont {Zantow}},\ }\href
  {\doibase 10.1016/S0370-2693(02)02415-2} {\bibfield  {journal} {\bibinfo
  {journal} {Phys. Lett. B}\ }\textbf {\bibinfo {volume} {543}},\ \bibinfo
  {pages} {41} (\bibinfo {year} {2002})},\ \Eprint
  {http://arxiv.org/abs/hep-lat/0207002} {arXiv:hep-lat/0207002} \BibitemShut
  {NoStop}%
\bibitem [{\citenamefont {Aoki}\ \emph {et~al.}(2009)\citenamefont {Aoki} \emph
  {et~al.}}]{Aoki:2009sc}%
  \BibitemOpen
  \bibfield  {author} {\bibinfo {author} {\bibfnamefont {Y.}~\bibnamefont
  {Aoki}} \emph {et~al.},\ }\href {\doibase 10.1088/1126-6708/2009/06/088}
  {\bibfield  {journal} {\bibinfo  {journal} {JHEP}\ }\textbf {\bibinfo
  {volume} {06}},\ \bibinfo {pages} {088} (\bibinfo {year} {2009})},\ \Eprint
  {http://arxiv.org/abs/0903.4155} {arXiv:0903.4155 [hep-lat]} \BibitemShut
  {NoStop}%
%%CITATION = 0903.4155;%%
\bibitem [{\citenamefont {Menezes}\ \emph
  {et~al.}(2009{\natexlab{b}})\citenamefont {Menezes}, \citenamefont
  {Benghi~Pinto}, \citenamefont {Avancini},\ and\ \citenamefont
  {Provid\^encia}}]{Menezes:2009uc}%
  \BibitemOpen
  \bibfield  {author} {\bibinfo {author} {\bibfnamefont {D.}~\bibnamefont
  {Menezes}}, \bibinfo {author} {\bibfnamefont {M.}~\bibnamefont
  {Benghi~Pinto}}, \bibinfo {author} {\bibfnamefont {S.}~\bibnamefont
  {Avancini}}, \ and\ \bibinfo {author} {\bibfnamefont {C.}~\bibnamefont
  {Provid\^encia}},\ }\href {\doibase 10.1103/PhysRevC.80.065805} {\bibfield
  {journal} {\bibinfo  {journal} {Phys.Rev.}\ }\textbf {\bibinfo {volume}
  {C80}},\ \bibinfo {pages} {065805} (\bibinfo {year} {2009}{\natexlab{b}})},\
  \Eprint {http://arxiv.org/abs/0907.2607} {arXiv:0907.2607 [nucl-th]}
  \BibitemShut {NoStop}%
%%CITATION = ARXIV:0907.2607;%%
\bibitem [{\citenamefont {Avancini}\ \emph {et~al.}(2011)\citenamefont
  {Avancini}, \citenamefont {Menezes},\ and\ \citenamefont
  {Provid\^encia}}]{Avancini:2011zz}%
  \BibitemOpen
  \bibfield  {author} {\bibinfo {author} {\bibfnamefont {S.~S.}\ \bibnamefont
  {Avancini}}, \bibinfo {author} {\bibfnamefont {D.~P.}\ \bibnamefont
  {Menezes}}, \ and\ \bibinfo {author} {\bibfnamefont {C.}~\bibnamefont
  {Provid\^encia}},\ }\href {\doibase 10.1103/PhysRevC.83.065805} {\bibfield
  {journal} {\bibinfo  {journal} {Phys. Rev.}\ }\textbf {\bibinfo {volume}
  {C83}},\ \bibinfo {pages} {065805} (\bibinfo {year} {2011})}\BibitemShut
  {NoStop}%
%%CITATION = PHRVA,C83,065805;%%
\bibitem [{\citenamefont {Avancini}\ \emph {et~al.}(2021)\citenamefont
  {Avancini}, \citenamefont {Farias}, \citenamefont {Pinto}, \citenamefont
  {Restrepo},\ and\ \citenamefont {Tavares}}]{Avancini:2020xqe}%
  \BibitemOpen
  \bibfield  {author} {\bibinfo {author} {\bibfnamefont {S.~S.}\ \bibnamefont
  {Avancini}}, \bibinfo {author} {\bibfnamefont {R.~L.~S.}\ \bibnamefont
  {Farias}}, \bibinfo {author} {\bibfnamefont {M.~B.}\ \bibnamefont {Pinto}},
  \bibinfo {author} {\bibfnamefont {T.~E.}\ \bibnamefont {Restrepo}}, \ and\
  \bibinfo {author} {\bibfnamefont {W.~R.}\ \bibnamefont {Tavares}},\ }\href
  {\doibase 10.1103/PhysRevD.103.056009} {\bibfield  {journal} {\bibinfo
  {journal} {Phys. Rev. D}\ }\textbf {\bibinfo {volume} {103}},\ \bibinfo
  {pages} {056009} (\bibinfo {year} {2021})},\ \Eprint
  {http://arxiv.org/abs/2008.10720} {arXiv:2008.10720 [hep-ph]} \BibitemShut
  {NoStop}%
\bibitem [{\citenamefont {Moreira}\ \emph {et~al.}(2012)\citenamefont
  {Moreira}, \citenamefont {Hiller}, \citenamefont {Osipov},\ and\
  \citenamefont {Blin}}]{Moreira:2010bx}%
  \BibitemOpen
  \bibfield  {author} {\bibinfo {author} {\bibfnamefont {J.}~\bibnamefont
  {Moreira}}, \bibinfo {author} {\bibfnamefont {B.}~\bibnamefont {Hiller}},
  \bibinfo {author} {\bibfnamefont {A.}~\bibnamefont {Osipov}}, \ and\ \bibinfo
  {author} {\bibfnamefont {A.}~\bibnamefont {Blin}},\ }\href {\doibase
  10.1142/S0217751X12500601} {\bibfield  {journal} {\bibinfo  {journal}
  {Int.J.Mod.Phys.}\ }\textbf {\bibinfo {volume} {A27}},\ \bibinfo {pages}
  {1250060} (\bibinfo {year} {2012})},\ \Eprint
  {http://arxiv.org/abs/1008.0569} {arXiv:1008.0569 [hep-ph]} \BibitemShut
  {NoStop}%
%%CITATION = ARXIV:1008.0569;%%
\bibitem [{\citenamefont {C\^amara~Pereira}(2016)}]{CamaraPereira:2016oxs}%
  \BibitemOpen
  \bibfield  {author} {\bibinfo {author} {\bibfnamefont {R.}~\bibnamefont
  {C\^amara~Pereira}},\ }\emph {\bibinfo {title} {{Chiral Transition and
  Deconfinement in Hybrid Stars}}},\ \href@noop {} {Master's thesis},\ \bibinfo
   {school} {Coimbra U.} (\bibinfo {year} {2016})\BibitemShut {NoStop}%
\end{thebibliography}%
\end{document}